\documentclass[10pt,journal,final,twocolumn,twoside]{IEEEtran}
\IEEEoverridecommandlockouts
\usepackage{amsmath,amssymb,amsfonts}
\usepackage{array}
\usepackage{cite}

\usepackage{graphicx}
\usepackage{textcomp}
\usepackage{xcolor}
\usepackage{times}
\usepackage{subfigure}  
\usepackage{acronym}  
\usepackage{balance}
\usepackage{threeparttable}
\usepackage{booktabs}

\usepackage{bm}
\usepackage{algorithm}
\usepackage{algorithmic}

\usepackage{lipsum}
\usepackage{stfloats}

\newtheorem{remark}{\bf Remark}

\usepackage{amsmath}
\allowdisplaybreaks[4]

\begin{document}
\title{Capacity Enhancement for Reconfigurable Intelligent Surface-Aided Wireless Network: \\from Regular Array to Irregular Array}

\author{{Ruochen Su,~\IEEEmembership{Student Member,~IEEE}, Linglong Dai,~\IEEEmembership{Fellow,~IEEE}, Jingbo Tan,~\IEEEmembership{Student Member,~IEEE}, Mo Hao, and Richard MacKenzie}

\thanks{Copyright (c) 2015 IEEE. Personal use of this material is permitted. However, permission to use this material for any other purposes must be obtained from the IEEE by sending a request to pubs-permissions@ieee.org. Part of this paper was presented at the IEEE GLOBECOM 2020~\cite{su2020capacity}.}
\thanks{R. Su, L. Dai, and J. Tan are with the Department of Electronic Engineering, Tsinghua University as well as the Beijing National Research Center for Information Science and Technology (BNRist), Beijing 100084, China (e-mails: src18@mails.tsinghua.edu.cn; daill@tsinghua.edu.cn; tanjb17@mails.tsinghua.edu.cn).}
\thanks{M. Hao is with the Tsinghua SEM Advanced ICT LAB, Tsinghua University, Beijing 100084, China (e-mail: haom@sem.tsinghua.edu.cn).}
\thanks{R. MacKenzie is with the BT TSO, Adastral Park, Ipswich, U.K. (e-mail: richard.mackenzie@bt.com).}
\thanks{This work was supported in part by the National Key Research and Development Program of China (Grant No. 2020YFB1807201), in part by	the National Natural Science Foundation of China (Grant No. 62031019), and in part by the European Commission through the H2020-MSCA-ITN META WIRELESS Research Project under Grant 956256. \textit{(Corresponding author:	Linglong Dai.)}}
\vspace{-2mm}
}

\maketitle
\begin{abstract}
Reconfigurable intelligent surface (RIS) is promising for future 6G wireless communications. However, the increased number of RIS elements results in the high overhead for channel acquisition and the non-negligible power consumption. Therefore, how to improve the system capacity with limited RIS elements is essential. Unlike the classical regular RIS whose elements are arranged on a regular grid, in this paper, we propose an irregular RIS structure to improve the system capacity. The key idea is to irregularly configure a given number of RIS elements on an enlarged surface, which provides extra spatial degrees of freedom compared with the regular RIS. In this way, the received signal power can be enhanced, and thus the system capacity can be improved. Then, we formulate a joint topology and precoding optimization problem to maximize the capacity for irregular RIS-aided communication systems. Accordingly, a joint optimization algorithm with low complexity is proposed to alternately optimize the RIS topology and the precoding design. Particularly, a tabu search-based method is used to design the irregular RIS topology, and a neighbor extraction-based cross-entropy method is introduced to optimize the precoding design. Simulation results demonstrate that, subject to the constraint of limited RIS elements, the proposed irregular RIS can significantly enhance the system capacity.
\end{abstract}

\begin{IEEEkeywords}
Reconfigurable intelligent surface (RIS), irregular design, topology optimization, precoding.
\end{IEEEkeywords}

\vspace{3mm}
\section{Introduction}\label{S1}
The wireless propagation environment between the base station (BS) and the user equipments (UEs) is generally regarded as uncontrollable. Thanks to the recent advances of metamaterials, the emerging reconfigurable intelligent surface (RIS) is able to control the propagation of incident signals via the interaction between electromagnetic waves and materials coated on the surface~\cite{pan2021reconfigurable}. Thus, the development of RIS makes it possible to manipulate the wireless propagation environment by tuning the reflection coefficients of a large number of RIS elements, which effectively improves the system capacity and reduces the communication outage~\cite{Ntontin'19,basar2019wireless,jung2019reliability}. To this end, RIS has been considered as a promising technology for 6G wireless communication systems~\cite{hu2018beyond,liang2019large}.

\subsection{Prior works}
As an emerging technology, RIS has recently attracted extensive attention including performance analysis~\cite{basar2019wireless}, precoding design~\cite{wang2020intelligent,huang2019reconfigurable,hu2021robust,yu2021irs,pan2020multicell,wu2019beamforming,liu2021compact}, channel state information acquisition~\cite{ma2021wideband,alwazani2020intelligent,mishra2019channel}, and hardware implementation~\cite{Dai2020reconfigurable}, etc. 

It has been revealed in~\cite{basar2019wireless} that the received signal power in RIS-aided wireless communication systems is quadratically proportional to the number of RIS elements. To achieve this exciting result, precoding design is essential for RIS-aided communication systems to jointly optimize the precoder at the BS and the reflection coefficients at the RIS. Different optimization metrics have been studied in recent works. Specifically, the sub-optimal analytical solution of precoding design for multiple RISs was derived in~\cite{wang2020intelligent}, which aimed to maximize the received signal power of a single-antenna user. The energy efficiency of a RIS-aided communication system was investigated by joint power allocation and precoding optimization for multi-user cases in~\cite{huang2019reconfigurable,yu2021irs}, which employed gradient descent search and sequential fractional programming. Moreover, the weighted sum-rate of all users in a RIS-aided communication system was maximized by utilizing the block coordinate descent algorithm, which alternately optimized the active precoding at the BS and the phase shifts at the RIS~\cite{pan2020multicell}. In addition, the transmit power at the BS was minimized by applying the successive refinement method~\cite{wu2019beamforming}. 

Although the precoding design of RIS-aided communication systems has been widely studied in the literature, prior works usually considered the classical regular RIS, where the RIS elements are regularly arranged on a grid with a constant inter-element spacing. Since the performance bound based on regular RIS highly relies on the number of RIS elements, the array gain provided by RIS can only be improved by increasing the number of RIS elements. However, the number of RIS elements is usually limited in practical systems, which is mainly caused by two reasons. The first one is the high overhead for the acquisition of channels. Specifically, the required pilot overhead of existing cascaded channel estimation schemes is usually proportional to the product of the number of RIS elements and the number of UEs \cite{ma2021wideband,alwazani2020intelligent}. Consequently, with a large number of RIS elements, the overhead for channel state information acquisition will be very high~\cite{hu2019two,basar2019wireless}. The second one is the limitation of the RIS power consumption. Most existing works have assumed that the power consumption of RIS elements is negligible. However, when the number of RIS elements is large, the RIS power consumption could be high. For example, the dissipated power of a RIS element is about 10 mW~\cite{huang2019reconfigurable}, so the total power consumption of RIS is up to 1 W with 100 RIS elements. Due to these two reasons, a not very large number of RIS elements is preferred in practice, which however limits the capacity of RIS-aided communication systems. Therefore, it is important to improve the capacity of RIS-aided communication systems with a limited number of RIS elements.

\subsection{Contributions}
In this paper, we focus on improving the capacity of RIS-aided communication systems with a limited number of RIS elements\footnote{Simulation codes are provided to reproduce the results in this paper: http://oa.ee.tsinghua.edu.cn/dailinglong/publications/publications.html.}. Unlike the existing regular RIS, we propose an irregular RIS structure to improve the system capacity. The contributions of this paper are summarized as follows.
\begin{itemize}{\tiny }
	\item We propose a new topological structure of RIS called irregular RIS, where a given number of elements are irregularly deployed on an enlarged surface. Considering that RIS is easy to be integrated into the wireless environment (such as the facades of buildings or ceilings)~\cite{basar2019wireless}, space available for the deployment of an enlarged surface is usually sufficient. To the best of our knowledge, the irregular RIS has not been investigated in the literature, which is studied in this paper for the first time. Since the selection of feasible locations for RIS elements can exploit spatial degrees of freedom (DoF), the proposed irregular RIS with an enlarged surface can enhance the system capacity.
	\item We develop the system model for the irregular RIS-aided communication systems, and formulate a joint topology and precoding optimization problem to improve the weighted sum-rate. Specifically, the topology design is formulated as an integer programming problem, while the precoding design is formulated to jointly optimize the precoder at the BS and the reflection coefficients at the irregular RIS.
	\item To solve the formulated optimization problem, we decouple the decision variables to be optimized and propose a joint optimization algorithm to alternately optimize the RIS topology and the precoding design. Particularly, an adaptive topology design method based on the tabu search algorithm~\cite{glover1989tabu} (ATS) is used to design the irregular RIS topology, and a neighbor extraction-based cross-entropy (NECE) precoding method is introduced for precoding design. We also analyze the complexity of the proposed joint optimization algorithm, which is much lower than that of the exhaustive search method. Finally, simulation results show that the proposed irregular RIS can significantly improve the capacity of RIS-aided communication systems. 
\end{itemize}

\subsection{Organization and notations}
\textit{Organization:} The rest of the paper is organized as follows. The model of the proposed irregular RIS-aided communication system is introduced in Section II. We formulate the weighted sum-rate optimization problem for the proposed system in Section III. A joint optimization algorithm is proposed to solve the formulated optimization problem in Section IV. Simulation results are shown in Section V, and the conclusions are finally drawn in Section VI.

\textit{Notations:} Vectors and matrices are denoted by lower-case and upper-case boldface letters, respectively. ${{{\bf{A}}^H}}$, ${{{\bf{A}}^T}}$, ${{{\bf{A}}^{ - 1}}}$, and ${{\bf{A}}(i,j)}$ denote the conjugate transpose, transpose, inverse, and ${(i,j)}$-th entry of matrix ${\bf{A}}$, respectively. ${{\left\|  \bf{a}  \right\|_1}}$ and ${{\left\|  \bf{a}  \right\|_2}}$ denote the ${{\ell}_1}$ norm and Euclidean norm of vector ${\bf{a}}$, respectively. ${{\rm{diag}}(\bf{a})}$ denotes a diagonal matrix whose diagonal elements consist of corresponding entries in vector ${\bf{a}}$. ${\left| {\cal A} \right|}$ denotes the cardinality of set ${{\cal A}}$. ${{{\mathbf{I}}_{K}}}$ denotes the identity matrix of size ${K \times K}$. Finally, ${{\mathbf{1}}_{K \times L}}$ denotes the all-one matrix of size ${K \times L}$.

\vspace{3mm}
\section{Irregular RIS-Aided Communication System}\label{S2}
In this section, we first introduce the concept of the proposed irregular RIS. Then, the model of the irregular RIS-aided communication system is illustrated.

\begin{figure*}[tp]
\begin{center}
\hspace*{0mm}\includegraphics[width=0.9\linewidth]{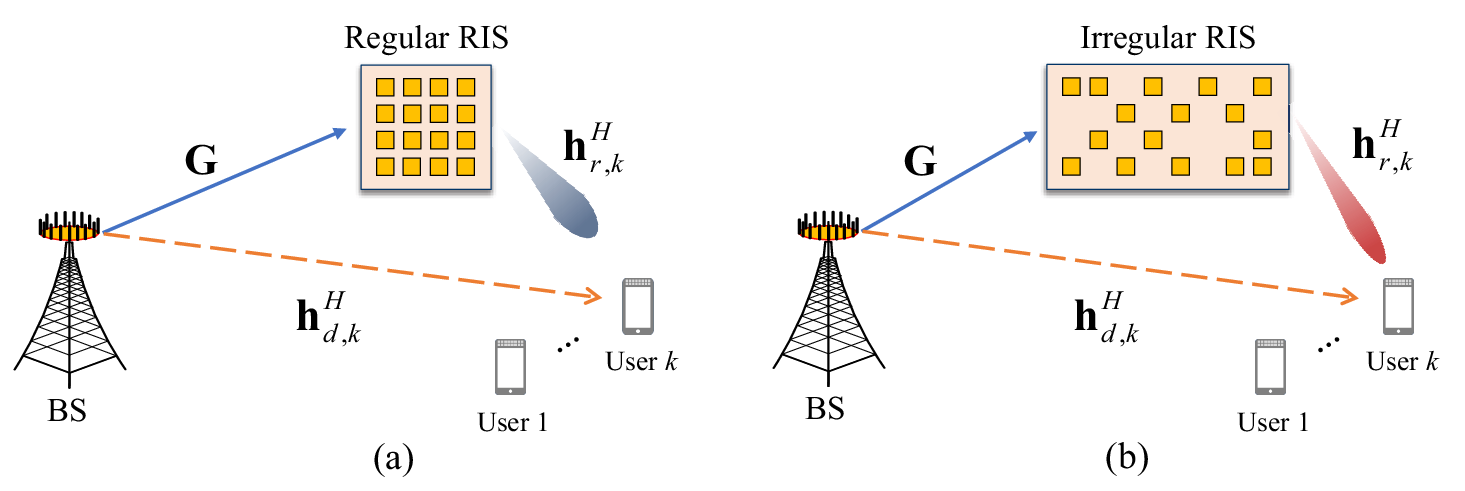}
\end{center}
\vspace*{-3mm}\caption{The RIS-aided communication system: (a) The classical regular RIS. (b) The proposed irregular RIS.} \label{FIG:architecture}
\end{figure*}

We consider a RIS-aided multiuser downlink multiple-input multiple-output (MIMO) wireless communication system in this paper. The existing regular RIS-aided communication system is shown in Fig. \ref{FIG:architecture} (a), where a BS equipped with ${M}$ antennas and a regular RIS composed of ${N}$ elements simultaneously serve ${K}$ single-antenna users. As shown in Fig. \ref{FIG:architecture} (b), different from the regular RIS whose elements are arranged on a regular surface with a constant inter-element spacing, we propose a new structure of irregular RIS, where ${N}$ RIS elements are irregularly distributed over ${N_{s}}$ grid points of an enlarged surface (${N_{s} \textgreater N}$). To simplify the description of the proposed concept, the grid constraint with a fixed grid spacing is assumed in this paper\footnote{The more complex scenario where elements can be arbitrarily distributed within the surface aperture is left for future works.}. Without loss of generality, the grid spacing between adjacent grid points is assumed to be half wavelength of the carrier frequency~\cite{wang2017analysis}. The proposed irregular RIS can be equivalently implemented by selecting feasible locations for RIS elements from all grid points. The unselected grid point can be switched off by connecting it to a matched load with a diode controller~\cite{rocca2016unconventional}. Besides, the tunable metasurface absorber can absorb the signal~\cite{liu2018programmable}, which provides another implementation to switch off the unselected grid points. It is worth noting that, the proposed irregular RIS is different from the classical irregular phased arrays, which aim to shape the beam pattern in the free space and suppress the sidelobe~\cite{rocca2016unconventional}. By contrast, the proposed irregular RIS working as a reflecting array aims to enhance the system capacity, which takes the influence of channel conditions and multiple UEs into account.

Let ${\mathbf{Z}={\rm{diag}}(\mathbf{z})}$ denote the topology matrix representing the topology of RIS, where ${\mathbf{z}=}$ ${[z_1,z_2,\cdots,z_{N_s}]^T}$. We define ${z_n \in \{1,0\}}$ to indicate whether a RIS element is deployed at the ${n}$-th grid point or not (${n=1,2,\cdots,N_s}$), i.e., ${z_n=1}$ means that the ${n}$-th grid point is selected for a RIS element, while ${z_n=0}$ denotes that the ${n}$-th grid point is not selected. 

The received signal ${\mathbf{y} \in \mathbb{C}^{K \times 1}}$ for all ${K}$ users can be expressed as
\begin{equation}\label{received signal}
\mathbf{y}=\left(\mathbf{H}_{r}^{H} \mathbf{Z} \mathbf{\Theta} \mathbf{G} + \mathbf{H}_{d}^{H}\right) \mathbf{x} + \mathbf{n},
\end{equation}
where ${\mathbf{x} \in \mathbb{C}^{M \times 1}}$ represents the transmitted signal at the BS, each component of ${\mathbf{n} \in \mathbb{C}^{K \times 1}}$ donates the additive white Gaussian noise (AWGN) with zero mean and variance ${\sigma^{2}}$, the reflection coefficients of ${N_s}$ grid points of the irregular RIS can be represented as
\begin{equation}\label{eq1-2}
\mathbf{\Theta}={\rm{diag}}([\beta_1 e^{j\theta_1},\beta_2 e^{j\theta_2},\cdots,\beta_{N_S} e^{j\theta_{N_s}}]),
\end{equation}
and ${\mathbf{G} \in \mathbb{C}^{N_s \times M}}$ denotes the BS-RIS channel. For simplifying the expression, we define ${\mathbf{H}_d^H =}$ ${\left[\mathbf{h}_{d,1},\mathbf{h}_{d,2},\cdots,\mathbf{h}_{d,K}\right]^H \in \mathbb{C}^{K \times M}}$, ${\mathbf{H}_r^H=\left[\mathbf{h}_{r,1},\mathbf{h}_{r,2},\cdots,\mathbf{h}_{r,K}\right]^H \in \mathbb{C}^{K \times N_s}}$, where ${\mathbf{h}_{d,k}^H}$ and ${\mathbf{h}_{r,k}^H}$ represent the channel from the BS to user ${k}$, and that from the RIS to user ${k}$, respectively. 

In this paper, we consider the fully digital precoder at the BS, i.e.,
\begin{equation}\label{eq1-3}
\mathbf{x}=\sum_{k=1}^{K} \mathbf{w}_{k} s_{k},
\end{equation}
where ${\mathbf{w}_k \in \mathbb{C}^{M \times 1}}$ denotes the precoding vector for user ${k}$, and ${\mathbf{s}= [s_1,s_2,\cdots,s_K]^H \in \mathbb{C}^{K \times 1}}$ denotes the transmitted symbol vector satisfying ${E[\mathbf{s} \mathbf{s}^H]=\mathbf{I}_{K}}$ for ${K}$ users. Considering the practical hardware implementation of RIS, we assume the constant reflection amplitude constraint and finite discrete phase shift constraint at the RIS~\cite{wu2019beamforming,Dai2020reconfigurable,tan2021hybrid,yang2016A}. To this end, for ${\forall n=1,2,\cdots,N_s}$, the reflection amplitude ${\beta_n}$ satisfies ${\beta_n=1}$, and the phase shift ${\theta_n}$ takes discrete values from the quantized phase shift set given by
\begin{equation}\label{set of phase shifts}
{\cal F}=\{ 0, \frac{2\pi}{2^b},\cdots,\frac{2\pi}{2^b}(2^b -1) \},
\end{equation}
where ${b}$ is the number of quantized bits of finite discrete phase shifts.

Based on the system model (\ref{received signal}), the signal-to-interference-plus-noise ratio (SINR) of user ${k}$ in the irregular RIS-aided communication system can be derived as 
\begin{equation}\label{SINR}
\gamma_{k}=\frac{\left|\left(\mathbf{h}_{r, k}^{H} \mathbf{Z} \mathbf{\Theta} \mathbf{G}+\mathbf{h}_{d, k}^{H}\right) \mathbf{w}_{k}\right|^{2}}{\sum_{i \neq k}^{K}\left|\left(\mathbf{h}_{r, k}^{H} \mathbf{Z} \mathbf{\Theta} \mathbf{G}+\mathbf{h}_{d, k}^{H}\right) \mathbf{w}_{i}\right|^{2}+\sigma^{2}}.
\end{equation}

Based on (\ref{SINR}), a joint topology and precoding optimization problem will be formulated for the proposed irregular RIS-aided communication system in the next section.

\vspace{3mm}
\section{Problem Formulation of the Proposed Irregular RIS-Aided Communication System}\label{S3}
To fully improve the system performance of the proposed irregular RIS-aided communication system, the topology design and the corresponding precoding design should be carefully devised. In this section, the joint topology and precoding design problem of the irregular RIS is firstly formulated to maximize the weighted sum-rate (WSR). Then, we convert the formulated problem to a precoding optimization problem for a specific topology of RIS by decoupling the decision variables.

Let ${\mathbf{W}=[{\mathbf{w}_1},{\mathbf{w}_2},\cdots,{\mathbf{w}_K}]}$ denote the fully digital precoder at the BS, ${\omega_k}$ denote the weight of user ${k}$. The transmit power at the BS is given by
\begin{equation}\label{eq4}
P_{\rm{T}}=\sum_{k=1}^{K} \left\|\mathbf{w}_k \right\|_2^2,
\end{equation}
which should be lower than the maximum allowable transmit power ${P}$. We focus on maximizing the WSR ${R}$ by jointly designing the irregular RIS topology and the corresponding precoding design, i.e., ${\mathbf{Z}}$, ${\mathbf{W}}$, and ${\mathbf{\Theta}}$. The WSR maximization problem can be formulated as
\begin{subequations}\label{eq5}
\begin{align}
\label{eq5a}
{\cal P}_{\rm{1}}: ~ & \max _{\mathbf{Z},\mathbf{W}, \mathbf{\Theta}} ~ R=\sum_{k=1}^{K} \omega_k \log _{2} \left(1+\gamma_{k}\right) \\
\label{eq5b}
& ~\, {\rm{s.t.}} ~~~~ C_1: P_{\rm{T}} \leq P, \\
\label{eq5c}
& ~~~~~~~~~\, C_2: \theta_{n} \in \mathcal{F}, \forall n=1,2,\cdots,N_{\rm{s}}, \\
\label{eq5d}
& ~~~~~~~~~\, C_3: z_n \in \{1,0\}, \forall n=1,2,\cdots,N_{\rm{s}}, \\
\label{eq5e}
& ~~~~~~~~~\, C_4: {\mathbf{1}}^T \mathbf{z}=N,
\end{align}
\end{subequations}
where ${C_1}$ denotes the transmit power constraint, ${C_2}$ denotes the discrete phase shift constraint, ${C_3}$ and ${C_4}$ denote the irregular RIS constraints. Note that ${C_3}$ and ${C_4}$ restrict the irregular topology design of RIS, where ${N}$ diagonal elements of the topology matrix ${\mathbf{Z}}$ are assigned with the value of ${1}$, and the rest ${N_s-N}$ diagonal elements are assigned with ${0}$.

Note that the objective function (\ref{eq5a}), the phase shift constraint ${C_2}$, and the irregular constraints ${C_3}$ and ${C_4}$ are non-convex. Thus, it is difficult to directly solve the WSR maximization problem ${{\cal P}_1}$. To tackle this challenge, we can decouple the RIS topology design and the corresponding precoding design. Specifically, for a given topology ${\mathbf{Z}_0}$, ${{\cal P}_1}$ can be reduced to
\begin{subequations}\label{eq6}
\begin{align}
\label{eq6a}
{\cal P}_{\rm{2}}: ~ & \max _{\mathbf{W}, \mathbf{\Theta}} ~ R=\sum_{k=1}^{K} \omega_k \log _{2} \left(1+\gamma_{k}\right) \\
\label{eq6b}
& \, {\rm{s.t.}} ~~~C_1: P_{\rm{T}} \leq P, \\
\label{eq6c}
& ~~~~~~~\, C_2: \theta_{n} \in \mathcal{F}, \forall n=1,2,\cdots,N_s, \\
\label{eq6d}
& ~~~~~~~\, C_3: \mathbf{Z}=\mathbf{Z}_0. 
\end{align}
\end{subequations}
The sub-problem ${{\cal P}_2}$ can be efficiently solved by alternating optimization, which will be discussed later in Section \ref{S4}.

\begin{remark}
The system model (\ref{received signal}) of the proposed irregular RIS-aided communication systems is reduced to the system model of the existing regular RIS-aided communication systems by simply assuming that
\begin{equation}\label{assumption}
N_s = N, \quad \mathbf{Z}={\mathbf{I}}_{N}.
\end{equation}
Thus, the formulated problem ${{\cal P}_1}$ can be seen as a general formulation for both regular RIS and irregular RIS. Similarly, if ${{\mathbf{Z}_0}={\mathbf{I}}_{N}}$, the problem ${{\cal P}_2}$ is reduced to ${{\cal P}_3}$ as
\begin{subequations}\label{eq6ex}
\begin{align}
\label{eq6exa}
{\cal P}_{\rm{3}}:~& \max _{\mathbf{W}, \mathbf{\Theta}} ~ R=\sum_{k=1}^{K} \omega_k \log _{2} \left(1+\gamma_{k}\right) \\
\label{eq6exb}
& \, {\rm{s.t.}} ~~~C_1: P_{\rm{T}} \leq P, \\
\label{eq6exc}
& ~~~~~~~\, C_2: \theta_{n} \in \mathcal{F}, \forall n=1,2,\cdots,N.
\end{align}
\end{subequations} 
We can find that the simplified problem ${{\cal P}_3}$ is just the precoding design problem in the existing regular RIS-aided communication systems~\cite{pan2020multicell}.
\end{remark}

\begin{figure}[tp]
	\begin{center}
		\hspace*{0mm}\includegraphics[width=0.9\linewidth]{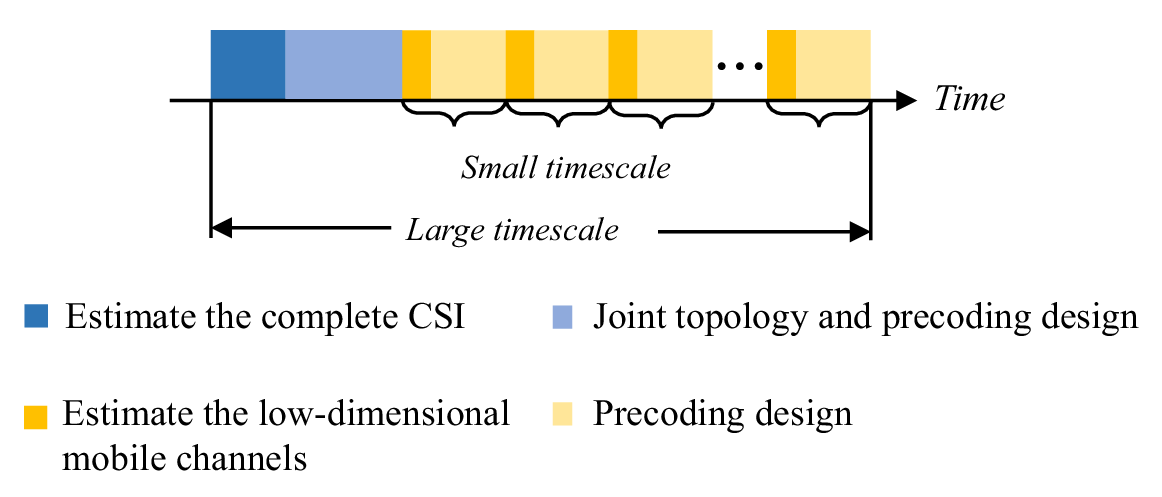}
	\end{center}
	\vspace*{-3mm}\caption{An illustration of the two-timescale irregular design framework.} \label{FIG:two-timescale}
\end{figure}

\begin{remark}
	Note that, the complete channel state information (CSI) of all grid points of RIS is required to solve ${{\cal P}_1}$. The CSI can be acquired by the classical methods, such as the least-squares estimation and the Bayesian minimum mean squared error estimation, which results in high pilot overhead~\cite{ma2021wideband,alwazani2020intelligent}. Fortunately, since the covariance matrix of the BS-RIS channel remains unchanged over a long period of time~\cite{hu2019two}, the designed topology of RIS can be temporarily fixed. Thus, we have proposed a two-timescale irregular design framework to reduce the pilot overhead and implementation complexity, as shown in Fig. \ref{FIG:two-timescale}. Specifically, the designed topology of RIS only needs to be changed adaptively in a large timescale, where the resultant overhead of estimating the complete CSI and the complexity of the joint topology and precoding design can be negligible. Then, the low-dimensional mobile channels for a given RIS topology, i.e., the BS-UE channel and the RIS-UE channel, are estimated in the small timescale, of which the required overhead is the same as that for a regular RIS with the same number of RIS elements. Accordingly, the precoding design with a given specific RIS topology, i.e., the solution to ${{\cal P}_2}$, is optimized in the small timescale to track the time-varying BS-UE channel and RIS-UE channel due to the user mobility. Therefore, from the perspective of a large timescale, the overhead and the complexity of the proposed two-timescale irregular design framework are almost the same as those of the channel acquisition and precoding of regular RIS with the same number of RIS elements.
\end{remark}

\vspace{3mm}
\section{The Proposed Joint Optimization Algorithm}\label{S4}
In this section, we propose a joint optimization algorithm to solve ${{\cal P}_1}$. First, an exhaustive search method is introduced in Subsection \ref{S4:s1}. Next, the overview of the proposed joint optimization algorithm is provided in Subsection \ref{S4:s2}. Specifically, we use an adaptive topology design method based on the tabu search algorithm (ATS) to optimize the irregular RIS topology in Subsection \ref{S4:s3}. Then, with a given specific RIS topology, the coupled variables ${\mathbf{W}}$ and ${\mathbf{\Theta}}$ in ${{\cal P}_2}$ are decoupled and solved by a neighbor extraction-based cross-entropy (NECE) precoding method in Subsection \ref{S4:s4}.

\subsection{Exhaustive search method}\label{S4:s1}
In this subsection, we first introduce an exhaustive search method to solve ${{\cal P}_1}$, which can be seen as a benchmark for all other possible solutions. Since the feasible sets of ${\mathbf{Z}}$ and ${\mathbf{\Theta}}$ are discrete sets with finite elements, the optimal solution can be obtained by the exhaustive search method.

The exhaustive search solution to ${{\cal P}_1}$ can be illustrated as follows. Firstly, all the irregular RIS topologies from ${\tbinom{N_s}{N}}$ possible options are traversed. Then, for each specific topology ${\mathbf{Z}}$, the phase shift for each RIS element is successively chosen to generate the reflection coefficients ${\mathbf{\Theta}}$ from ${2^{bN}}$ possible candidates. Subsequently, given ${\mathbf{Z}}$ and ${\mathbf{\Theta}}$, the precoder ${\mathbf{W}}$ can be obtained by the zero-forcing (ZF) based linear precoder\footnote{The precoding at the BS is not the focus of this paper. Without loss of generality, the ZF based linear precoder at the BS can also be replaced by the minimum mean squared error (MMSE) based linear precoder.}~\cite{zhao2022rethinking}. The equivalent channel between the BS and UEs is given by
\begin{equation}\label{equivalent channel}
\mathbf{H}_{\rm{eq}}=\mathbf{H}_{r}^{H} \mathbf{Z} \mathbf{\Theta} \mathbf{G} + \mathbf{H}_{d}^{H}.
\end{equation}
Thus, the ZF precoder can be expressed as~\cite{wu2019beamforming}
\begin{equation}\label{eq7}
\mathbf{W}=\mathbf{H}_{\rm{eq}}^H (\mathbf{H}_{\rm{eq}} \mathbf{H}_{\rm{eq}}^H)^{-1}\mathbf{P}_{\rm{B}}^{\frac{1}{2}},
\end{equation}
where ${\mathbf{P}_{\rm{B}}}$ represents the power allocation at the BS. Through the exhaustive search method, we calculate the WSR ${R}$ for all different combinations of ${\mathbf{Z}}$ and ${\mathbf{\Theta}}$, from which the optimal solution will be finally found. However, the possible number of searches of the exhaustive search method is up to ${\tbinom{N_s}{N} \times 2^{bN}}$, which is prohibitively high for a large RIS size.

\subsection{Overview of the proposed joint optimization algorithm}\label{S4:s2}
\begin{figure}[tp]
\begin{center}
\hspace*{0mm}\includegraphics[width=0.9\linewidth]{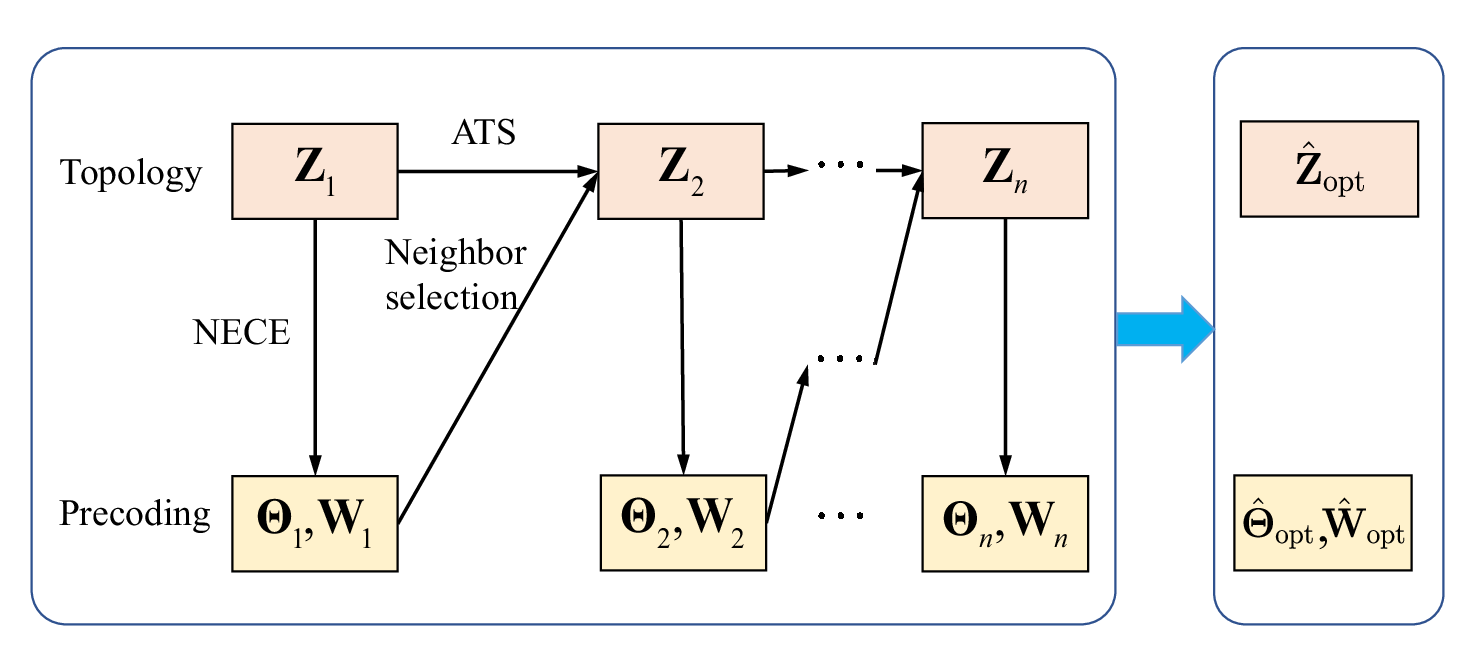}
\end{center}
\vspace*{-3mm}\caption{The proposed joint optimization algorithm.} \label{FIG:framework}
\end{figure}

Due to the intolerable complexity of the exhaustive search method, we propose a sub-optimal joint optimization algorithm with low complexity to solve ${{\cal P}_1}$. As shown in Fig. \ref{FIG:framework}, we decouple the decision variables and alternately optimize the RIS topology and the precoding design. 

Specifically, instead of traversing all possible topological structures of RIS, we search for the sub-optimal design of RIS iteratively by an ATS method. Specifically, in each iteration, the problem ${{\cal P}_1}$ is reduced to ${{\cal P}_2}$, where the RIS topology ${\mathbf{Z}}$ is fixed. We generate the neighbors of ${\mathbf{Z}}$ and obtain corresponding precoding designs by a NECE method with ZF precoder at the BS. By comparing the WSR of these neighbors, the optimal neighbor is selected for the next iteration. When the maximum number of iterations is reached or the termination condition is satisfied, we obtain the sub-optimal solution to ${{\cal P}_1}$, which is denoted by ${\hat{\mathbf{Z}}_\text{opt}}$, ${\hat{\mathbf{\Theta}}_\text{opt}}$, and ${\hat{\mathbf{W}}_\text{opt}}$.

The details of topology design of the irregular RIS will be provided in the following Subsection \ref{S4:s3}. The precoding design of the irregular RIS-aided communication system will be provided in Subsection \ref{S4:s4}. The generalization of our proposed joint optimization algorithm will be discussed in Subsection \ref{S4:s5}. The complexity analysis of the proposed joint optimization algorithm will be analyzed in Subsection \ref{S4:s6}.

\subsection{ATS-based irregular topology design}\label{S4:s3}
Inspired by the tabu search algorithm in~\cite{glover1989tabu}, we propose an ATS method to optimize the irregular topology of RIS, which obtains a sub-optimal RIS topoloy by iteratively searching the possible topologies according to an adaptive move criterion.

The proposed ATS method is summarized in {\bf Algorithm \ref{alg:1}}, and its explanations are detailed as follows. Firstly we initialize the RIS topology ${\mathbf{Z}_1}$ by randomly selecting ${N}$ ones and ${N_s-N}$ zeros in step 1. For a given ${\mathbf{Z}_1}$, the WSR can be calculated based on ZF precoder at the BS and discrete phase shift adjustment at the RIS in step 2, which will be discussed in the next subsection. In step 4, we focus on the neighbors of ${\mathbf{Z}_i}$ generated by a redefined move criterion in the ${i}$-th iteration, where we randomly swap ${p}$ ones for ${p}$ zeros among the diagonal elements of ${\mathbf{Z}_i}$. Here a neighbor is defined as a new topology matrix with the different RIS topology, and ${p}$ is defined as the neighbor distance which will be dynamically adjusted according to the value of ${i}$. We can choose a large ${p}$ at the beginning of iterations for a wide searching range. As the iteration progresses, we decrease the neighbor distance ${p}$ for fine tuning. Next, the obtained neighbors should be checked in the tabu list, which avoids one topology being searched more than once. Those tabu solutions will be discarded in step 5. In this way, we obtain ${Q}$ neighbors and separately calculate the corresponding WSR for all ${Q}$ candidates ${\{{\mathbf{Z}}_i^q\}_{q=1}^{Q}}$ in step 6. Then, we select the best candidate with the maximum WSR, which is saved as the new topology ${\mathbf{Z}_{i+1}}$ for the next iteration. The optimal ${\hat{\mathbf{Z}}_{\text{opt}}}$ and ${\hat{R}_{\text{opt}}}$ are updated accordingly in step 7. The tabu list is updated in step 8-11. Specifically, the feasible solution ${\mathbf{Z}_i}$ is added to the tabu list to avoid cyclic search and the earliest tabu solution in the current list is removed. After the iteration threshold ${I_{\rm{T}}}$ is reached, the sub-optimal topology of RIS can be obtained.

\begin{algorithm}[t]
	\renewcommand{\algorithmicrequire}{ \textbf{Inputs:}} 
	\renewcommand{\algorithmicensure}{ \textbf{Output:}} 
	\caption{$\!\!${\bf :} Adaptive topology design method based on the tabu search algorithm.}
	\label{alg:1}
	\begin{algorithmic}[1]
		\REQUIRE ~
		Tabu list ${{\cal H}={\emptyset}}$, storage size ${H_{\rm{size}}}$, neighbor distance ${p}$, neighborhood size ${Q}$, and number of iterations ${I_\text{T}}$.
		\ENSURE ~
		RIS topology ${\hat{\mathbf{Z}}_\text{opt}}$.
		\STATE Initialize the RIS topology ${\mathbf{Z}_1}$ by randomly selecting ${N}$ ones and ${N_s-N}$ zeros.
		\STATE Calculate the WSR ${R_1}$ by a NECE precoding method based on the initial topology ${\mathbf{Z}_1}$ and save ${\hat{\mathbf{Z}}_{\text{opt}}=\mathbf{Z}_1}$, ${\hat{R}_{\text{opt}}=R_1}$.
		\FOR{${i=1:I_\text{T}}$}
		\STATE Update ${p}$ according to the value of ${i}$. Randomly swap ${p}$ ones for ${p}$ zeros among the diagonal elements of ${\mathbf{Z}_i}$ and generate alternative neighbors of ${\mathbf{Z}_i}$. 
		\STATE Discard the tabu neighbors in ${\cal H}$ and obtain ${Q}$ candidates ${\{{\mathbf{Z}}_i^q\}_{q=1}^{Q}}$.
		\STATE Calculate the WSR ${R}$ for all the candidates.
		\STATE Select the candidate with the maximum ${R}$ as the new topology ${\mathbf{Z}_{i+1}}$. Update the optimal ${\hat{\mathbf{Z}}_{\text{opt}}}$ and ${\hat{R}_{\text{opt}}}$.
		\IF{${|{\cal H}| < H_\text{size}}$}
		\STATE Add ${\mathbf{Z}_i}$ into the tabu list ${\cal H}$.
		\ELSE
		\STATE Add ${\mathbf{Z}_i}$ to the end of the tabu list ${\cal H}$ and remove the head of ${\cal H}$.
		\ENDIF
		\ENDFOR
	\end{algorithmic}
\end{algorithm}

Different from the tabu search algorithm developed in~\cite{glover1989tabu}, the neighbor, move criterion, and objective function of the proposed ATS method are redefined accordingly in terms of the irregular topology design of RIS. Besides, the neighbor distance ${p}$ in the move criterion is adaptively adjusted based on the iteration process. A large value of ${p}$ corresponds to a wide searching range, while a smaller one aims at fine tuning and better convergence.

\subsection{NECE-based precoding design}\label{S4:s4}
The use of the ATS method requires joint precoding at the BS and the RIS for a specific RIS topology, which corresponds to solving ${{\cal P}_2}$. The successive refinement (SR) method proposed in~\cite{wu2019beamforming} only optimizes one of the ${N}$ phase shifts in each iteration by fixing the others, which is likely to fall into local optimal. In this paper, we propose a NECE method to solve ${{\cal P}_2}$, which iteratively optimizes the precoder at the BS and the phase shifts at the RIS with probability distribution updated during the iterative process.

The proposed NECE precoding method is summarized in {\bf Algorithm \ref{alg:2}}, and further explanations are detailed as follows. Let ${\mathbf{P}=[\mathbf{p}_1,\mathbf{p}_2,\cdots,\mathbf{p}_{N_s}]}$ denote the probability matrix of ${\mathbf{\Theta}}$ for a given ${\mathbf{Z}}$, where ${\mathbf{p}_n \in \mathbb{R}^{2^b \times 1}}$ is the probability vector of ${\theta_n}$ satisfying ${\left\|\mathbf{p}_{n}\right\|_1=1}$. Each component of ${\mathbf{p}_{n}}$ denotes the probability of taking different values from ${\cal F}$ as shown in (\ref{set of phase shifts}). Let ${{\cal F}(k)}$ denote the ${k}$-th element of ${\cal F}$. We assign an indicator function expressed as
\begin{equation}\label{delta function}
	\delta(t)=\left\{\begin{array}{lc}
		1, & \text {t = 0,} \\
		0, & \text {t $\neq$ 0.}
	\end{array}\right.
\end{equation}

\begin{algorithm}[t]
	\caption{$\!\!${\bf :} Neighbor extraction-based cross-entropy precoding method}
	\renewcommand{\algorithmicrequire}{ \textbf{Inputs:}} 
	\renewcommand{\algorithmicensure}{ \textbf{Outputs:}} 
	\label{alg:2}
	\begin{algorithmic}[1]
		\REQUIRE ~
		RIS topology ${\mathbf{Z}}$, BS-RIS channel ${\mathbf{G}}$, RIS-UE channel ${\mathbf{H}_r}$, BS-UE channel ${\mathbf{H}_d}$, number of iterations ${I_\text{N}}$, number of candidates ${C}$, number of primary elites ${C_{\rm{pr}}}$, and quantized phase shift set ${\cal F}$.
		\ENSURE ~
		Reflection coefficients ${\hat{\mathbf{\Theta}}_\text{opt}}$ and precoder ${\hat{\mathbf{W}}_\text{opt}}$.
		\STATE Initialize the probability matrix ${\mathbf{P}^{(1)}=\frac{1}{2^b}\times \mathbf{1}_{2^b \times N_s}}$.
		\FOR{${i=1:I_\text{N}}$}
		\STATE Randomly generate ${C}$ candidates ${\{{\mathbf{\Theta}}^c\}_{c=1}^{C}}$ based on the PDF ${\Xi(\mathbf{\Theta};\mathbf{P}^{(i)})}$.
		\STATE Calculate the effective channel ${\mathbf{H}_{\rm{eq}}^c}$.
		\STATE Calculate the precoders ${\{ \mathbf{W}^c \}_{c=1}^{C}}$.
		\STATE Calculate the WSR ${\{R({\mathbf{\Theta}}^c)\}_{c=1}^{C}}$.
		\STATE Sort ${\{R({\mathbf{\Theta}}^c)\}_{c=1}^{C}}$ in an descending order as \\
		${R({\mathbf{\Theta}}^{[1]}) \geq R({\mathbf{\Theta}}^{[2]}) \geq \cdots \geq R({\mathbf{\Theta}}^{[C]}) }$.
		\STATE Select ${C_\text{pr}}$ best candidates as primary elites \\
		${{\mathbf{\Theta}}^{[1]},{\mathbf{\Theta}}^{[2]},\cdots,{\mathbf{\Theta}}^{[C_\text{pr}]}}$.
		\STATE Obtain ${N(2^b-1)}$ extra candidates through neighbor extraction based on ${{\mathbf{\Theta}}^{[1]}}$.
		\STATE Select ${C_\text{sup}}$ candidates with better performance than ${\mathbf{\Theta}}^{[1]}$ as supplementary elites. 
		\STATE Update ${C_\text{elite}=C_\text{pr}+C_\text{sup}}$ and calculate the weight ${\eta_c}$ of each elite ${\mathbf{\Theta}}^{[c]}$, ${c=1,2,\cdots,C_\text{elite}}$.
		\STATE Update ${\mathbf{P}^{(i+1)}}$ based on ${\{ \mathbf{\Theta}^{[c]} \}_{c=1}^{C_\text{elite}}}$ and ${\{ \eta_{c} \}_{c=1}^{C_\text{elite}}}$.
		\ENDFOR
	\end{algorithmic}
\end{algorithm}

Firstly, we initialize the probability matrix in step 1 given by
\begin{equation}\label{probability matrix}
\mathbf{P}^{(1)}=\frac{1}{2^b}\times \mathbf{1}_{2^b \times N_s},
\end{equation}
which represents that the value of ${\theta_n}$ is selected from the elements in ${\cal F}$ with the same probability. In step 3, we randomly generate ${C}$ candidates ${\{{\mathbf{\Theta}}^c\}_{c=1}^{C}}$ based on the probability distribution function (PDF) ${\Xi(\mathbf{\Theta};\mathbf{P}^{(i)})}$ in the ${i}$-th iteration, which can be expressed as
\begin{equation}\label{eq8}
\Xi(\mathbf{\Theta};\mathbf{P}^{(i)})=\prod_{n=1}^{N_s}\left(\prod_{k=1}^{2^b}(p_{n,k}^{(i)})^{\delta(\theta_n-{\cal F} (k))} \right).
\end{equation}
Note that ${\mathbf{\Theta}}^{i}$ is the generated ${i}$-th candidate. Then, we are able to calculate the effective channel ${\mathbf{H}_\text{eq}^c}$ for all candidates ${\{{\mathbf{\Theta}}^c\}_{c=1}^{C}}$ according to (\ref{equivalent channel}) in step 4. The precoders ${\{ \mathbf{W}^c \}_{c=1}^{C}}$ and WSR ${\{R({\mathbf{\Theta}}^c)\}_{c=1}^{C}}$ can be calculated in step 5-6 according to (\ref{eq7}) and (\ref{eq5a}), respectively. Next, we sort ${\{R({\mathbf{\Theta}}^c)\}_{c=1}^{C}}$ in a descending order in step 7 as
\begin{equation}\label{sort}
	{R({\mathbf{\Theta}}^{[1]}) \geq R({\mathbf{\Theta}}^{[2]}) \geq \cdots \geq R({\mathbf{\Theta}}^{[C]}) },
\end{equation}
where ${\mathbf{\Theta}}^{[i]}$ represents the ${i}$-th candidate in the descending sequence, from which ${C_\text{pr}}$ best candidates are selected as primary elites in step 8. 

Then, we propose a neighbor extraction method to expand the searching range by changing the phase shift of each diagonal element of ${{\mathbf{\Theta}}^{[1]}}$. Notice that only ${N}$ diagonal elements of ${{\mathbf{\Theta}}^{[1]}}$ are effective since the trace of ${\mathbf{Z}}$ is ${N}$ according to (\ref{eq5e}), whose values can be selected from ${\cal F}$. In this way, we can obtain ${N(2^b-1)}$ extra candidates in step 9 and calculate the corresponding precoders at the BS as well as the WSR. The candidate whose WSR is larger than ${R({\mathbf{\Theta}}^{[1]})}$ is selected as a supplementary elite in step 10. The total number of elites is updated as ${C_\text{elite}=C_\text{pr}+C_\text{sup}}$, where ${C_\text{sup}}$ is the number of selected supplementary elites. The weight ${\eta_{c}}$ of the ${c}$-th elite is calculated in step 11 as follows:
\begin{equation}\label{elite weight}
	\eta_{c}= \frac{R({\mathbf{\Theta}}^{[c]}) C_{\text{elite}}} {\sum_{c=1}^{C_{\text{elite}}} R(\mathbf{\Theta}^{[c]})} ,
\end{equation}
which represents the ratio of the WSR of elite ${c}$ to the average WSR of all elites. Note that the larger WSR corresponds to the larger weight. After that, in step 12, we update ${\mathbf{P}^{(i+1)}}$ according to the probability transfer criterion, i.e.,
\begin{equation}\label{probability}
	\mathbf{P}^{(i+1)}=\arg \max _{\mathbf{P}^{(i)}} \frac{1}{C_\text{elite}} \sum_{c=1}^{C_{\text {elite }}} \eta_{c} \ln \Xi\left(\mathbf{\Theta}^{[c]} ; \mathbf{P}^{(i)}\right).
\end{equation}
The updated probability matrix ${\mathbf{P}^{(i+1)}}$ is employed in the next loop until the iteration threshold ${I_{\rm{N}}}$ is reached. Finally, the sub-optimal precoding design can be obtained.

The contributions of the NECE method proposed in this paper are summarized as follows. Different from the cross-entropy algorithm developed in~\cite{rubinstein2013cross}, the probability vector and objective function of the NECE method are redefined in terms of the specific problem, which has a greater probability to avoid falling into local optimal. In addition, neighbor extraction and individual weights of elites are introduced for expanded searching range and improved accuracy.

So far we have introduced the details of the proposed joint optimization algorithm. The generalization and the complexity will be discussed in the following subsections.

\subsection{Generalization of the proposed joint optimization algorithm}\label{S4:s5}
The proposed joint optimization algorithm can be generalized in various optimization scenarios. For example, the transmit power minimization problem is discussed in this subsection, which provides another perspective to evaluate our proposed schemes. 

Let ${\gamma_{k}}$ denote the practical SINR of user ${k}$ as expressed in (\ref{SINR}), and ${\mu_k}$ denote the target SINR of user ${k}$. The transmit power minimization problem can be described as
\begin{subequations}\label{eq13}
	\begin{align}
		\label{eq13a}
		{\cal P}_{4}: ~ & \min _{\mathbf{Z},\mathbf{W}, \mathbf{\Theta}} ~ P_{\text{T}}=\sum_{k=1}^{K} \left\|\mathbf{w}_{k}\right\|_2^{2} \\
		\label{eq13b}
		& ~\, {\rm{s.t.}} ~~~~ C_1: \gamma_{k} \geq \mu_{k}, \forall k=1,2,\ldots,K, \\
		\label{eq13c}
		& ~~~~~~~~~\, C_2: \theta_{n} \in \mathcal{F}, \forall n=1,2,\cdots,N_\text{s}, \\
		\label{eq13d}
		& ~~~~~~~~~\, C_3: z_n(z_n-1)=0, \forall n=1,2,\cdots,N_\text{s}, \\
		\label{eq13e}
		& ~~~~~~~~~\, C_4: {\mathbf{1}}^T \mathbf{z}=N.
	\end{align}
\end{subequations}

${{\cal P}_4}$ can also be solved by the proposed joint optimization algorithm through sightly modifying the optimization target. Similar to the procedure to solve ${{\cal P}_1}$, we search for the sub-optimal design of RIS topology iteratively by the ATS method. In each iteration, with a fixed RIS topology ${\mathbf{Z}}$, we generate the neighbors of ${\mathbf{Z}}$ and obtain the corresponding precoding designs by the NECE method. By comparing the required transmit power, the optimal neighbor is selected for the next iteration. Finally, we can obtain the sub-optimal solution to ${{\cal P}_4}$.

\subsection{Complexity analysis}\label{S4:s6}
In this subsection, we quantitatively compare the searching complexity, i.e., the maximum number of searches, of the exhaustive search method and the proposed joint optimization algorithm.

In the exhaustive search method, ${\binom{N_s}{N}}$ possible RIS topologies are traversed with ${N}$ elements deployed over ${N_s}$ grid points. For a specific topology, the possible combinations of phase shifts at the RIS with ${N}$ elements is up to ${2^{bN}}$, where ${b}$ is the quantized bit number. Thus, the searching complexity of the optimal solution is 
\begin{equation}\label{eq10}
C_\text{opt}=\binom{N_s}{N} \times 2^{bN}.
\end{equation}

By contrast, we only search ${I_\text{T} Q}$ possible RIS topologies in the ATS method, where ${I_\text{T}}$ is the required number of iterations for convergence, and ${Q}$ represents the neighborhood size. As for the NECE precoding method, given a specific ${\mathbf{Z}}$, ${I_\text{N} (C+N(2^{b}-1))}$ combinations are tested, where ${I_\text{N}}$ is the number of iterations of the NECE method, ${C}$ and ${N(2^{b}-1)}$ denote the number of primary candidates and extra candidates in each iteration, respectively. Therefore, the searching complexity of the proposed joint optimization algorithm is
\begin{equation}\label{eq11}
C_\text{p}=Q I_\text{T} I_\text{N} (C+N(2^{b}-1)).
\end{equation}
Notice that ${I_\text{T}}$ and ${I_\text{N}}$ of the proposed algorithms are usually small, which will be confirmed by simulations in Section \ref{S5}. 

The comparison of the searching complexity between our proposed algorithm and the exhaustive search method is shown in Table I. The typical parameters are set to ${Q=15}$, ${I_\text{T}=40}$, ${I_\text{N}=15}$, ${C=200}$, and ${b=1}$ for a large-scale system with ${N_{\rm{s}} = 64}$ and ${N = 32}$, which can guarantee the convergence of our proposed algorithm in simulations. One can observe that the searching complexity of the proposed algorithm is much lower than that of the exhaustive search method. Actually, as the system size increases, the searching complexity of the exhaustive search method becomes unacceptable. However, the searching complexity of the proposed algorithm is just linear with ${N}$, which is much more computationally efficient. 

\begin{table}[htp]
\setlength{\abovecaptionskip}{-5pt}
\setlength{\belowcaptionskip}{0pt}
\caption{The Comparison of Searching Complexity} \label{TAB1}
\begin{center}
\begin{threeparttable}
\begin{tabular}{lcc}
\toprule[1pt]
 & Exhaustive search method & Proposed algorithm \\
\hline \\ [-2 ex]
 ${N_{\rm{s}} = 16, N = 8}$ & 3294720 & 1872000 \\
 ${N_{\rm{s}} = 64, N = 32}$ & ${7.8711\times10^{27}}$ & 2088000 \\
\toprule[1pt]
\end{tabular}
\end{threeparttable}
\end{center}
\end{table}

\vspace{3mm}
\section{Simulation Results}\label{S5}
In this section, we provide simulation results to evaluate the performance of the proposed irregular RIS-aided communication systems by employing the proposed joint optimization algorithm.

\subsection{Simulation setup}\label{S5:s1}
\begin{figure}[tp]
\begin{center}
\hspace*{0mm}\includegraphics[width=0.9\linewidth]{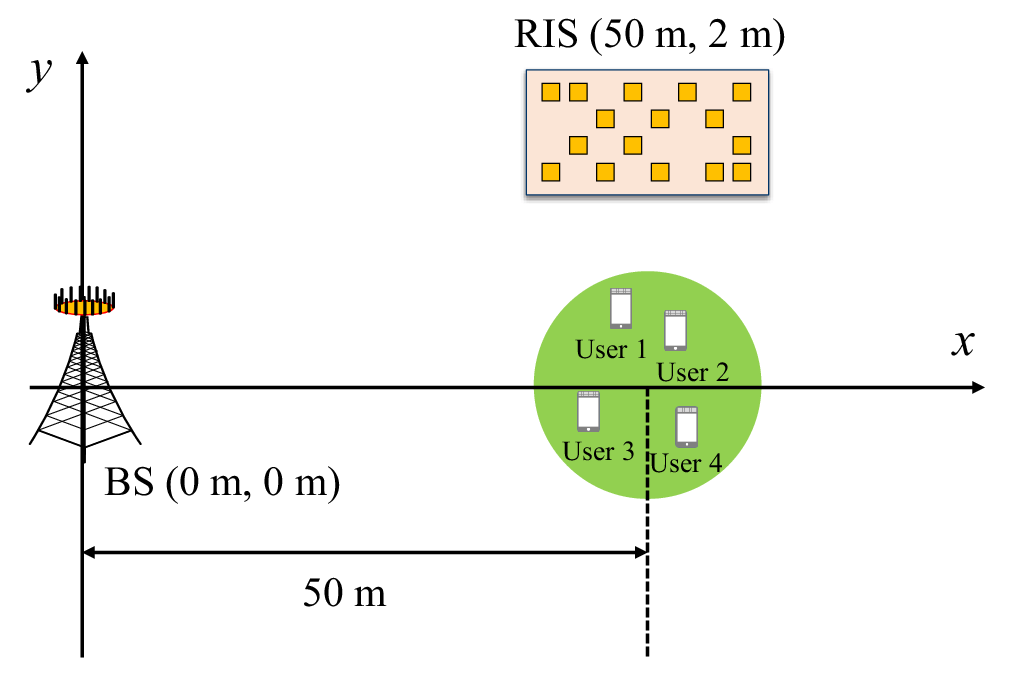}
\end{center}
\vspace*{-3mm}\caption{The simulation scenario of the proposed irregular RIS-aided communication system.} \label{FIG:scenario}
\end{figure}

As shown in Fig. \ref{FIG:scenario}, we consider an irregular RIS-aided multiuser MIMO communication system for simulations. The BS equipped with ${M}$ antennas is located at ${(0 \ \text{m}, 0 \ \text{m})}$. The irregular RIS with ${N}$ elements distributed over a rectangular surface with ${N_s}$ grid points is located at ${(50 \ \text{m}, 2 \ \text{m})}$. ${K}$ single-antenna users are randomly scattered in a circular area, whose center is located at ${(50 \ \text{m}, 0 \ \text{m})}$.

The Rician fading channel model is adopted to account for the small-scale fading. Therefore, the BS-RIS channel can be given by
\begin{equation}\label{Rician fading}
	\mathbf{G} = \sqrt{\frac{\beta_{\rm{BR}}}{1+\beta_{\rm{BR}}}} \mathbf{G}^{\rm{LoS}} + \sqrt{\frac{1}{1+\beta_{\rm{BR}}}} \mathbf{G}^{\rm{NLoS}},
\end{equation}
where ${\mathbf{G}^{\rm{LoS}}}$ denotes the LoS component, ${\mathbf{G}^{\rm{NLoS}}}$ denotes the NLoS component, i.e., the uncorrelated Rayleigh fading component, and ${\beta_{\rm{BR}}}$ denotes the Rician factor. The BS-UE channel and the RIS-UE channel can be obtained similarly. The large-scale fading, i.e., the distance-dependent path loss, is considered as well. Specifically, the path loss of the BS-RIS-UE channel can be expressed as~\cite{Ozdogan'19,zhang2020joint}
\begin{equation}\label{eq12}
f_{\rm{r}}(d_\text{BR},d_\text{RU})=C_{\rm{r}}d_{\rm{BR}}^{-\alpha_{\rm{BR}}} d_{\rm{RU}}^{-\alpha_{\rm{RU}}},
\end{equation}
where ${d_\text{BR}}$ and ${d_\text{RU}}$ are the distance between the BS and RIS, and that between the RIS and UEs, respectively. ${C_{\rm{r}}}$ denotes the effect of channel fading and antenna gain. ${\alpha_{\rm{BR}}}$ and ${\alpha_{\rm{RU}}}$ denote the path loss exponent. Similarly, the path loss of the BS-UE channel is expressed as ${f_{\rm{d}}(d_\text{BU})=C_{\rm{d}}d_{\rm{BU}}^{-\alpha_{\rm{BU}}}}$. The parameters are set to ${C_{\rm{r}}=-60}$ dB, ${C_{\rm{d}}=-30}$ dB, ${\alpha_{\rm{BR}}=2}$, ${\alpha_{\rm{RU}}=2}$, ${\alpha_{\rm{BU}}=3.5}$, ${\beta_{\rm{BR}}=0}$, ${\beta_{\rm{BU}}=0}$, and ${\beta_{\rm{RU}}=1}$~\cite{zhang2020joint}.

Considering that the 1-bit phase shift for metasurfaces is relatively easy to implement with diodes~\cite{yang2016A}, we set the quantized bit number as ${b=1}$. Besides, the noise power is set to ${\sigma^2=-80}$ dBm, the weights of users are set to ${\omega_k=1, \forall k=1,2,\cdots,K}$. The parameters of proposed iteration algorithms are set to ${Q=15}$, ${I_{\rm{T}}=40}$, ${I_{\rm{N}}=15}$, ${C=200}$, ${C_{\rm{pr}}=40}$, if not particularly indicated. The neighbor distance ${p}$ depends on the irregular ratio of RIS, i.e., ${\frac{N}{N_s}}$, and is dynamically changing during the iteration. For instance, if ${N=32}$ and ${N_s=64}$, ${p}$ is set to 3 at the beginning and is reduced to 2 when the iteration reaches half of the maximum number of iterations. Due to the excessive possible cases of RIS topologies, the size of the tabu list can be simply set to ${1}$ to expand the searching range, with a very low probability of falling into a loop. To validate the superiority of the proposed irregular RIS and joint optimization algorithm, we assume that the CSI is perfectly known~\cite{wang2020intelligent,huang2019reconfigurable,pan2020multicell,wu2019beamforming,liu2021compact}.

\subsection{WSR performance}
In this subsection, we illustrate the WSR performance of the proposed irregular RIS-aided communication system. The solution to ${{\cal P}_3}$, i.e., the WSR optimization problem of the classical regular RIS-aided communication systems\cite{pan2020multicell}, is used as a benchmark, which serves as the baseline. The proposed joint optimization algorithm is used to solve the formulated problem ${{\cal P}_1}$. Specifically, the ATS method is utilized in the irregular topology design of RIS. Besides, the SR method is adopted to optimize the phase shifts of RIS for comparison with the NECE precoding method proposed in this paper~\cite{wu2019beamforming}.
\begin{figure}[tp]
\begin{center}
\hspace*{0mm}\includegraphics[width=0.9\linewidth]{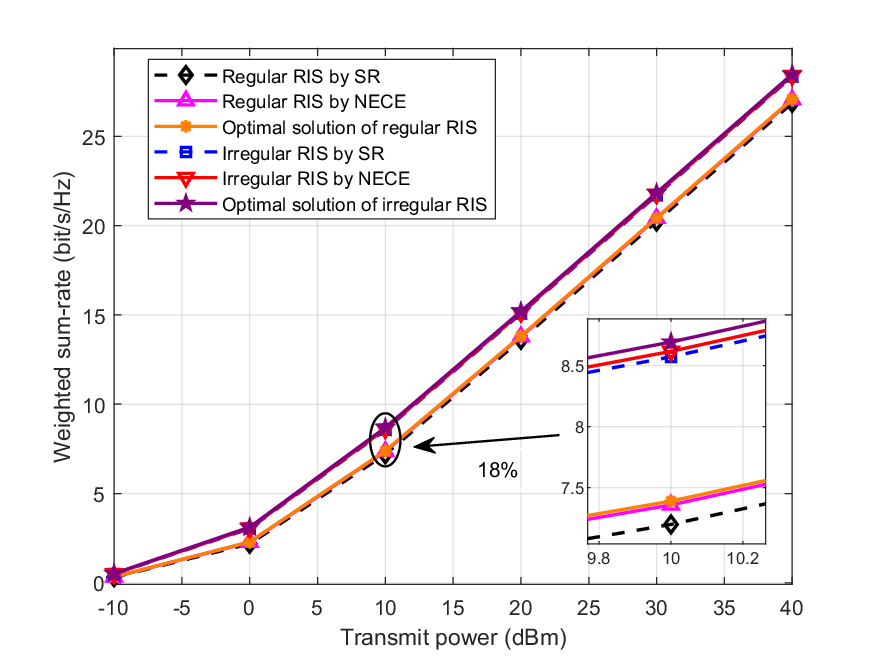}
\end{center}
\vspace*{-3mm}\caption{WSR versus the transmit power. ${M=4}$, ${N=8}$, ${N_s=16}$, ${K=2}$.} \label{FIG:simulation_small}
\end{figure}

We first consider a small-scale system with ${M=4}$, ${N=8}$, ${N_s=16}$, and ${K=2}$. The optimal solution to ${{\cal P}_1}$ based on the exhaustive search method serves as the upper bound. The WSR versus the transmit power for the small size system is shown in Fig. \ref{FIG:simulation_small}. By comparing the upper bounds, one can observe that the proposed irregular RIS-based scheme with a limited number of RIS elements can enhance the system capacity compared with the existing solution. For example, the WSR of the proposed scheme increases by ${18 \%}$ at the transmit power of ${10}$ dBm. Besides, note that both the NECE method and the SR method are effective, we can conclude that the capacity enhancement of the proposed irregular RIS-aided communication system does not depend on a specific precoding algorithm. The reason is that, the selection of feasible locations for RIS element deployment leads to additional spatial DoF, which enables us to select an ${N}$-element subset with the optimal channel conditions. Therefore, the irregular RIS with ${N}$ elements can achieve the spatial diversity benefits of ${N_s}$ grid points, which enhances the received signal power and thus improves the system capacity via the well-designed topology and precoding design. It also implies that the proposed joint optimization algorithm based on ATS method and NECE method is able to achieve the sub-optimal performance with much lower complexity, which verifies the reliability and convergence of our proposed joint optimization algorithm.

\begin{figure}[tp]
	\begin{center}
		\hspace*{0mm}\includegraphics[width=0.9\linewidth]{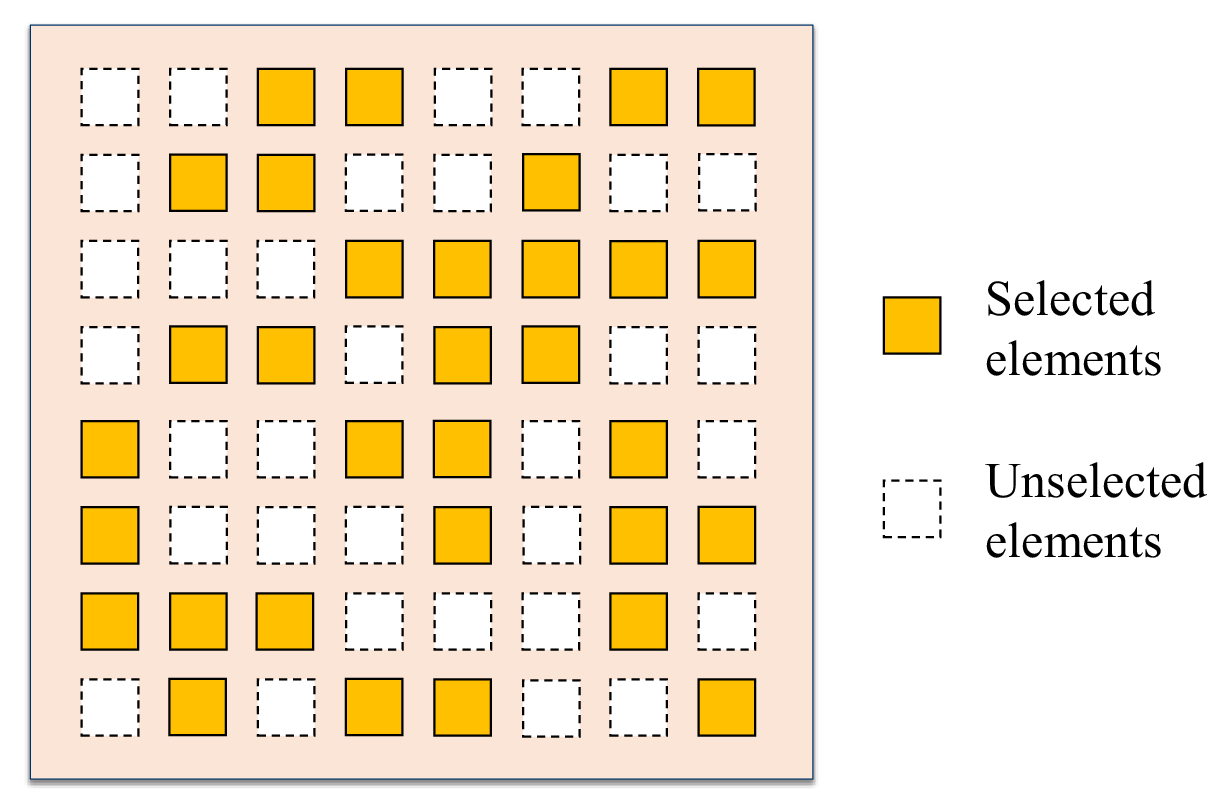}
	\end{center}
	\vspace*{-3mm}\caption{The optimized RIS topology. ${N=32}$, ${N_s=64}$.} \label{FIG:simulation_topology}
\end{figure}
\begin{figure}[tp]
\begin{center}
\hspace*{0mm}\includegraphics[width=0.9\linewidth]{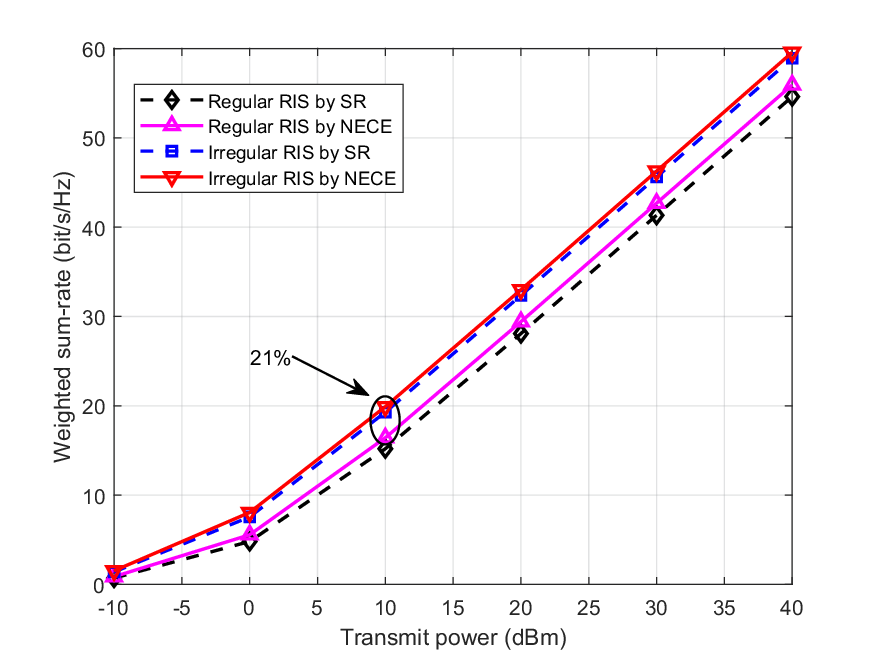}
\end{center}
\vspace*{-3mm}\caption{WSR versus the transmit power. ${M=4}$, ${N=32}$, ${N_s=64}$, ${K=4}$.} \label{FIG:simulation_large}
\vspace*{-0.5em}
\end{figure}

Then, we consider a large-scale system with ${M=4}$, ${N=32}$, ${N_s=64}$, and ${K=4}$. Since the complexity of the exhaustive search method is unacceptable for such a large system size, we only provide the sub-optimal solution by the proposed joint optimization algorithm. Here we provide an optimized RIS topology in Fig. \ref{FIG:simulation_topology}, where the colored squares represent the selected locations for RIS elements. The precoding optimization for the classical regular RIS-aided communication systems based on the NECE method is utilized as a benchmark scheme. Besides, the SR method for both cases of the classical regular RIS and the proposed irregular RIS is realized for comparison. The WSR versus the transmit power is shown in Fig. \ref{FIG:simulation_large}, where similar conclusions can be drawn. The WSR of the proposed irregular RIS is improved compared with that of the regular RIS. For instance, provided the NECE method is adopted, the WSR of the proposed irregular scheme increases by ${21 \%}$ at the transmit power of ${10}$ dBm. In addition, the proposed NECE method outperforms the SR method in both cases of the existing regular RIS and the proposed irregular RIS, which confirms the effectiveness of the proposed NECE method.

\subsection{Irregular ratio of RIS}
\begin{figure}[tp]
	\begin{center}
		\hspace*{0mm}\includegraphics[width=0.9\linewidth]{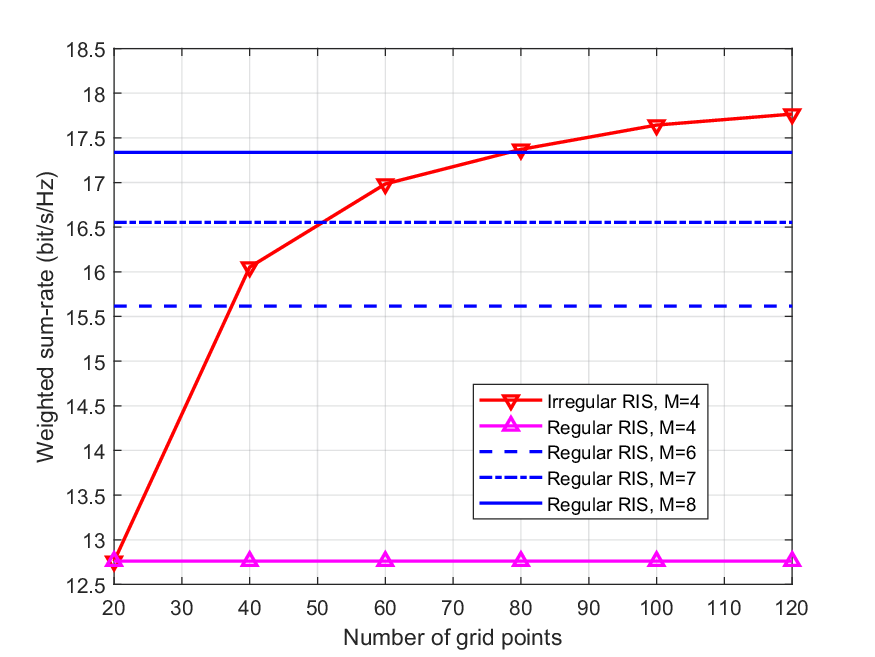}
	\end{center}
	\vspace*{-3mm}\caption{WSR versus the number of grid points of the irregular RIS. ${P_{\rm{T}}=10}$ dBm, ${N=20}$, ${K=4}$.} \label{FIG:simulation_size}
\end{figure}

The effect of different irregular ratios of RIS is analyzed in this subsection. First we consider an irregular RIS with a fixed number of elements, where ${M=4}$, ${N=20}$, and ${K=4}$. The size of the irregular RIS is variable, which is represented by ${N_s}$, i.e., the number of grid points of the surface. The simulation results are provided based on the proposed joint optimization algorithm with the transmit power set to ${P_{\rm{T}}=10}$ dBm. The existing regular RIS-based scheme with ${M=4}$, ${N=20}$, and ${K=4}$ serves as a benchmark, whose surface size is fixed. The WSR versus the number of grid points of the irregular RIS is shown in Fig. \ref{FIG:simulation_size}. It is observed that altering the irregular ratio of the irregular RIS via enlarging the surface size effectively improves the WSR performance, where the number of RIS elements is assumed to be a constant. Moreover, we consider several regular RIS-based scenarios where more antennas are deployed at the BS, namely, ${M=6}$, ${M=7}$, and ${M=8}$. The number of RIS elements and the number of users remain unchanged. It implies that the irregular RISs with enlarged size, namely, ${N_s=40}$, ${N_s=60}$, and ${N_s=80}$, outperform the classical regular RISs with ${M=6}$, ${M=7}$, and ${M=8}$, respectively. Note that when the irregular ratio is ${25\%}$, i.e., ${N=20}$ and ${N_s=80}$, the irregular scheme saves half of the number of antennas at the BS. Thus, the irregular RIS can provide a feasible solution to improve the system performance without increasing antennas and RF chains at the BS, whose hardware cost and power consumption are usually high. Nevertheless, with the increased number of grid points of the irregular RIS, the growth of performance slows down and the system complexity increases. Thus, we should make a tradeoff between the cost and the performance by carefully designing the irregular ratio of RIS in practice.

\begin{figure}[tp]
	\begin{center}
		\hspace*{0mm}\includegraphics[width=0.9\linewidth]{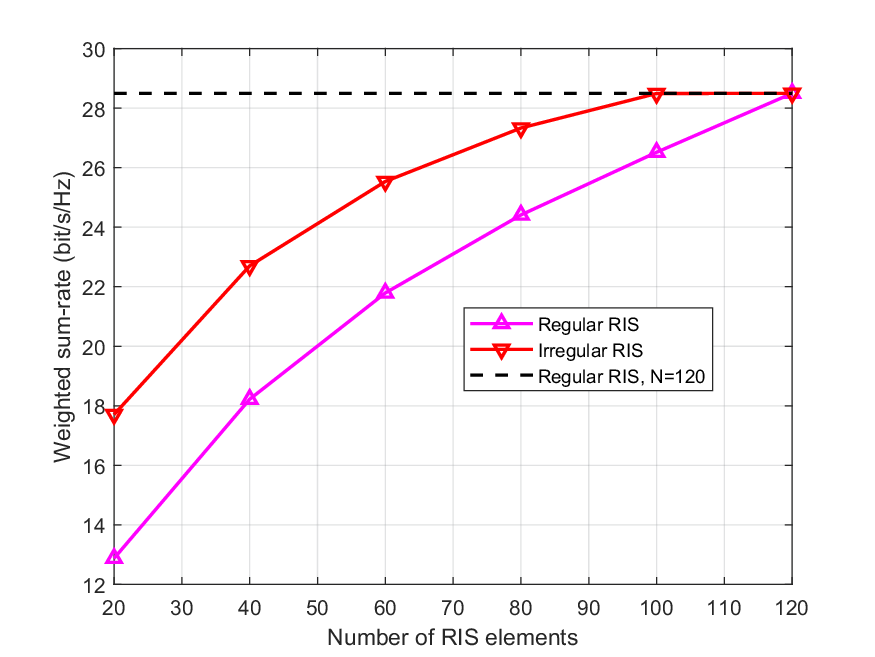}
	\end{center}
	\vspace*{-3mm}\caption{WSR versus the number of RIS elements. ${P_{\rm{T}}=10}$ dBm, ${N_s=120}$, ${K=4}$.} \label{FIG:simulation_elements}
\end{figure}
Then, we consider a RIS with variable elements ${N}$ and a fixed size. The transmit power is set to ${P_{\rm{T}}=10}$ dBm. Other parameters are set to ${M=4}$, ${N_s=120}$, and ${K=4}$. The WSR versus the number of RIS elements is shown in Fig. \ref{FIG:simulation_elements}, where the classical regular RIS-based scheme with ${M=4}$, ${K=4}$, and different numbers of RIS elements serves as the benchmark scheme. One can observe that, the proposed irregular RIS-based scheme with ${N_s=120}$ achieves higher WSR than the classical regular scheme. Besides, the red curve illustrates the scalability of our proposed irregular RIS-based scheme. Specifically, the system capacity increases with the increased number of RIS elements. Therefore, it implies that we can design a RIS with a large aperture and the number of elements selected for communication depends on the capacity requirements. It is worth noting that the performance gap between the proposed irregular RIS and the regular RIS with $N=120$ becomes smaller as the irregular ratio of RIS increases, hence the benefit of further increasing RIS elements with a very high irregular ratio is negligible. The tradeoff between the cost and performance can be a possible future research topic.

Furthermore, we analyze the pilot overhead for channel estimation and the RIS power consumption of the proposed irregular RIS-based scheme and the classical regular RIS-based scheme. It is assumed that there are $\alpha$ small timescales in a large timescale in the proposed two-timescale irregular design framework. The minimum required pilot overhead for a regular RIS with $N$ elements in a small timescale can be given by $\frac{2(N+1)}{\alpha}+K\left\lceil\frac{N}{M}\right\rceil+K$~\cite{hu2019two}. Similarly, the minimum required pilot overhead for an irregular RIS with $N$ elements and $N_s$ grid points can be given by $\frac{2(N_s+1)}{\alpha}+K\left\lceil\frac{N}{M}\right\rceil+K$. Besides, the dissipated power of a RIS element is about 10 mW~\cite{huang2019reconfigurable}. Thus, considering a case where the parameters are set to $M=4$, ${N=60}$, $N_s=120$, $K=4$, and $\alpha=30$, the comparison between the irregular scheme and the regular schemes is provided in Table II. One can observe that the proposed irregular RIS with ${N=60}$ and ${N_s=120}$ saves 45\% of the pilot overhead and 50\% of the power consumption compared with the regular RIS with ${N=120}$ elements. In addition, the pilot overhead and the power consumption of the irregular scheme are almost the same as those of the regular scheme with ${N=60}$ RIS elements. Thus, we can conclude that the proposed irregular RIS provides a better tradeoff between the performance and the complexity as well as the cost, which can enhance the system capacity with a limited number of RIS elements.
\begin{table}[htp]
	\setlength{\abovecaptionskip}{-5pt}
	\setlength{\belowcaptionskip}{0pt}
	\caption{The Comparison of Pilot Overhead and Power Consumption} \label{TAB2}
	\begin{center}
		\begin{threeparttable}
				\begin{tabular}{lcc}
					\toprule[1pt]
					& Pilot overhead & Power consumption \\
					\hline \\ [-2 ex]
					Regular RIS with ${N=60}$ & 68 & 0.6 W \\
					Irregular RIS & 72 & 0.6 W \\
					Regular RIS with ${N=120}$ & 132 & 1.2 W \\
					\toprule[1pt]
				\end{tabular}
		\end{threeparttable}
	\end{center}
\end{table}

\begin{figure}[tp]
	\begin{center}
		\hspace*{0mm}\includegraphics[width=0.9\linewidth]{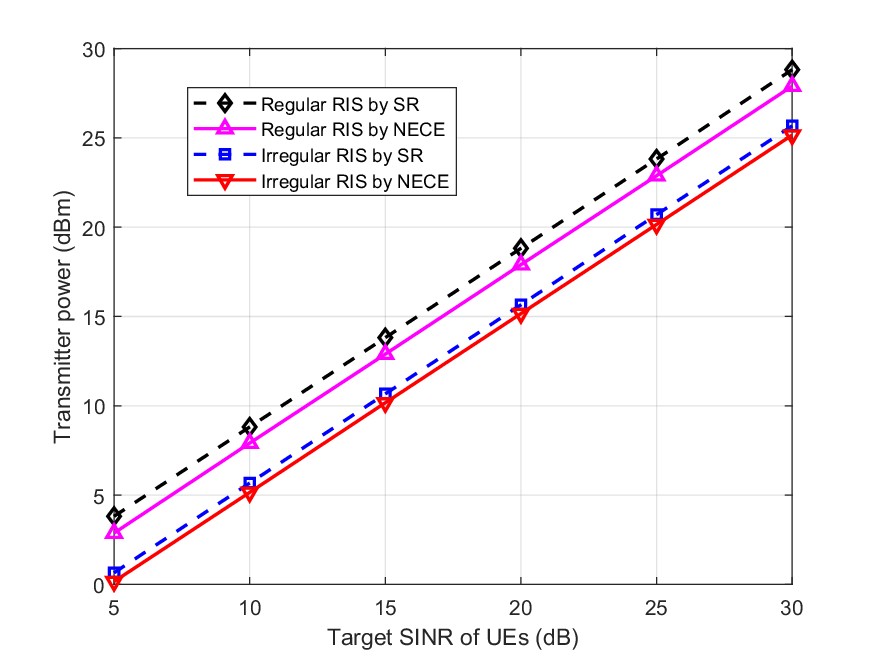}
	\end{center}
	\vspace*{-3mm}\caption{Transmit power versus the target SINR of UEs. ${M=4}$, ${N=32}$, ${N_s=64}$, ${K=4}$.} \label{FIG:simulation_powerReduce}
\end{figure}

\subsection{Transmit power consumption}
The solution of the transmit power minimization problem ${{\cal P}_4}$ is analyzed in this subsection. We show the transmit power versus the target SINR of UEs in Fig. \ref{FIG:simulation_powerReduce} with parameters set to ${M=4}$, ${N=32}$, ${N_s=64}$ and ${K=4}$. It is revealed that the proposed irregular scheme requires much lower transmit power than the classical one for a certain target SINR. In addition, the proposed NECE method outperforms the SR method for both schemes. Thus, we can draw the conclusion that the irregular scheme can significantly decrease the transmit power consumption, which further verifies the effectiveness of the proposed irregular RIS and the corresponding joint optimization algorithm.

\subsection{Time-varying channels}
\begin{figure}[tp]
	\begin{center}
		\hspace*{0mm}\includegraphics[width=0.9\linewidth]{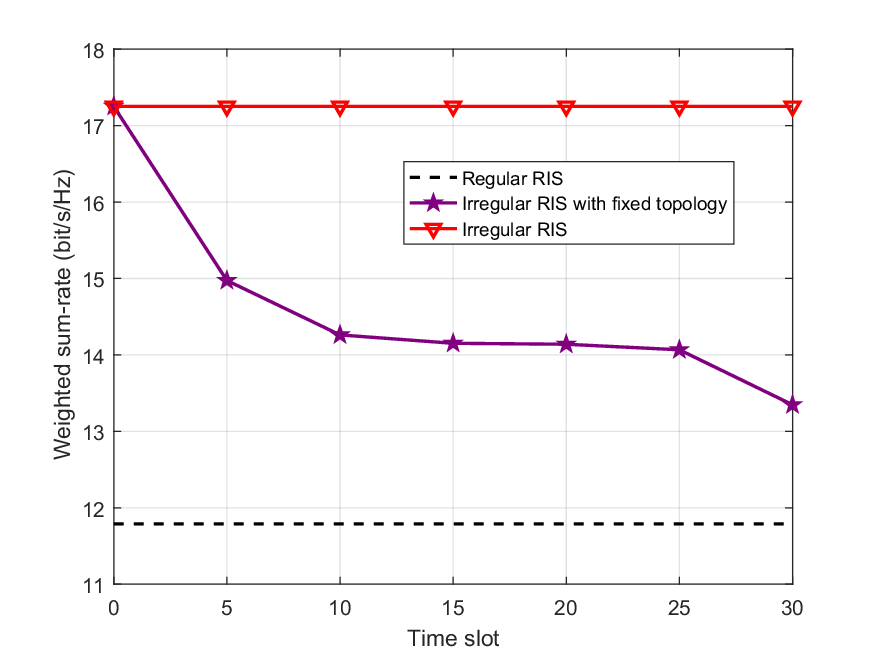}
	\end{center}
	\vspace*{-3mm}\caption{WSR versus the sample time with time-varying channels. ${M=4}$, ${N=32}$, ${N_s=64}$, ${K=4}$.} \label{FIG:simulation_timevarying}
\end{figure}
The performance of the proposed two-timescale irregular design framework with time-varying channels is evaluated in this subsection. ${K=4}$ single-antenna users are served by the BS equipped with ${M=4}$ antennas. The regular RIS is composed of ${N=32}$ elements. The irregular RIS is composed of ${N=32}$ elements distributed over ${N_s=64}$ grid points. The transmit power is set to 10 dBm. The BS-RIS channel is generated by the Rician channel in Section \ref{S5:s1}. Since the covariance matrix of the BS-RIS channel remains unchanged in a large timescale, we assume that the BS-RIS channel is changed at the beginning of each large timescale. To illustrate the mobility of UEs, we adopt the ``CDL-B'' time-varying channel model to describe the BS-UE channel and the RIS-UE channel in a large timescale with 30 time slots. The length of a time slot is 0.625 ms, which represents a small timescale. The velocity of each UE is 3 km/h. The number of channel paths is 23. The carrier frequency is 28 GHz. The following cases are considered in a large timescale. Case 1: The regular RIS. Case 2: The irregular RIS with the fixed topology in the large timescale. Case 3: The irregular RIS. The WSR versus the time slot with time-varying channels are shown in Fig. \ref{FIG:simulation_timevarying}. One can observe that the performance of Case 3 serves as the upper bound, which optimizes the topology of RIS in each time slot. For Case 2, the designed topology of RIS is fixed in the large timescale. The WSR of Case 2 decreases slowly with the time slot changes, whose decreasing trend tends to be gradual. The final time slot of Case 2 is in the next large timescale with totally changed channels, where a steep drop of WSR occurs. Besides, the WSR of Case 2 always outperforms that of Case 1. Therefore, the proposed two-timescale irregular design framework is confirmed to significantly reduce the overhead and complexity with acceptable performance loss.

\vspace{3mm}
\section{Conclusions}\label{S6}
The capacity of the existing regular RIS-aided wireless communication systems with a limited number of RIS elements is restricted. To tackle this challenge, we investigated the design of irregular RIS in this paper. Firstly, we proposed an irregular RIS structure with a given number of elements distributed over an enlarged surface. Then, for the proposed irregular RIS-aided communication system, we formulated a WSR maximization problem to optimize the system capacity. Finally, a joint optimization algorithm with low complexity was proposed to iteratively solve the optimization problem. Specifically, an ATS method was used to design the irregular RIS topology, and a NECE method was introduced to optimize the precoding design. Simulation results validated that, with a limited number of RIS elements, the proposed irregular RIS can enhance the system capacity. Several open problems are left for future works. For example, how to deploy the RIS elements arbitrarily within the surface aperture of other possible shapes~\cite{rocca2016unconventional} remains to be investigated. Moreover, other metrics such as the energy efficiency of irregular RIS-aided communication systems are worth further investigations.

\bibliographystyle{IEEEtran} 
\bibliography{IEEEabrv,SuSparseRIS}

\begin{thebibliography}{10}
\providecommand{\url}[1]{#1}
\csname url@samestyle\endcsname
\providecommand{\newblock}{\relax}
\providecommand{\bibinfo}[2]{#2}
\providecommand{\BIBentrySTDinterwordspacing}{\spaceskip=0pt\relax}
\providecommand{\BIBentryALTinterwordstretchfactor}{4}
\providecommand{\BIBentryALTinterwordspacing}{\spaceskip=\fontdimen2\font plus
\BIBentryALTinterwordstretchfactor\fontdimen3\font minus
  \fontdimen4\font\relax}
\providecommand{\BIBforeignlanguage}[2]{{%
\expandafter\ifx\csname l@#1\endcsname\relax
\typeout{** WARNING: IEEEtran.bst: No hyphenation pattern has been}%
\typeout{** loaded for the language `#1'. Using the pattern for}%
\typeout{** the default language instead.}%
\else
\language=\csname l@#1\endcsname
\fi
#2}}
\providecommand{\BIBdecl}{\relax}
\BIBdecl

\bibitem{su2020capacity}
R.~Su, L.~Dai, J.~Tan, M.~Hao, and R.~MacKenzie, ``Capacity enhancement for
  irregular reconfigurable intelligent surface-aided wireless communications,''
  in \emph{Proc. IEEE Global Commun. Conf. (IEEE GLOBECOM’20)}, Taipei, Dec.
  2020, pp. 1--6.

\bibitem{pan2021reconfigurable}
C.~Pan, H.~Ren, K.~Wang, J.~F. Kolb, M.~Elkashlan, M.~Chen, M.~Di~Renzo,
  Y.~Hao, J.~Wang, A.~L. Swindlehurst, X.~You, and L.~Hanzo, ``Reconfigurable
  intelligent surfaces for {6G} systems: {Principles}, applications, and
  research directions,'' \emph{IEEE Commun. Mag.}, vol.~59, no.~6, pp. 14--20,
  Jun. 2021.

\bibitem{Ntontin'19}
M.~D. Renzo, K.~Ntontin, J.~Song, F.~H. Danufane, X.~Qian, F.~Lazarakis, J.~D.
  Rosny, D.~Phan-Huy, O.~Simeone, R.~Zhang, M.~Debbah, G.~Lerosey, M.~Fink,
  S.~Tretyakov, and S.~Shamai, ``Reconfigurable intelligent surfaces vs.
  relaying: Differences, similarities, and performance comparison,'' \emph{IEEE
  Open J. Commun. Soc.}, vol.~1, pp. 798--807, Jun. 2020.

\bibitem{basar2019wireless}
E.~Basar, M.~Di~Renzo, J.~De~Rosny, M.~Debbah, M.-S. Alouini, and R.~Zhang,
  ``Wireless communications through reconfigurable intelligent surfaces,''
  \emph{IEEE Access}, vol.~7, pp. 116\,753--116\,773, Aug. 2019.

\bibitem{jung2019reliability}
M.~Jung, W.~Saad, Y.~Jang, G.~Kong, and S.~Choi, ``Reliability analysis of
  large intelligent surfaces ({LISs}): Rate distribution and outage
  probability,'' \emph{IEEE Wireless Commun. Lett.}, vol.~8, no.~6, pp.
  1662--1666, Dec. 2019.

\bibitem{hu2018beyond}
S.~Hu, F.~Rusek, and O.~Edfors, ``Beyond massive {MIMO}: {The} potential of
  data transmission with large intelligent surfaces,'' \emph{IEEE Trans. Signal
  Process.}, vol.~66, no.~10, pp. 2746--2758, Oct. 2018.

\bibitem{liang2019large}
Y.-C. Liang, R.~Long, Q.~Zhang, J.~Chen, H.~V. Cheng, and H.~Guo, ``Large
  intelligent surface/antennas ({LISA}): Making reflective radios smart,''
  \emph{J. Commun. Inf. Netw.}, vol.~4, no.~2, pp. 40--50, Jun. 2019.

\bibitem{wang2020intelligent}
P.~Wang, J.~Fang, X.~Yuan, Z.~Chen, and H.~Li, ``Intelligent reflecting
  surface-assisted millimeter wave communications: Joint active and passive
  precoding design,'' \emph{IEEE Trans. Veh. Technol.}, vol.~69, no.~12, pp.
  14\,960--14\,973, Dec. 2020.

\bibitem{huang2019reconfigurable}
C.~Huang, A.~Zappone, G.~C. Alexandropoulos, M.~Debbah, and C.~Yuen,
  ``Reconfigurable intelligent surfaces for energy efficiency in wireless
  communication,'' \emph{{IEEE} Trans. Wireless Commun.}, vol.~18, no.~8, pp.
  4157--4170, Aug. 2019.

\bibitem{hu2021robust}
S.~Hu, Z.~Wei, Y.~Cai, C.~Liu, D.~W.~K. Ng, and J.~Yuan, ``Robust and secure
  sum-rate maximization for multiuser {MISO} downlink systems with
  self-sustainable {IRS},'' \emph{IEEE Trans. Commun.}, vol.~69, no.~10, pp.
  7032--7049, Oct. 2021.

\bibitem{yu2021irs}
X.~Yu, D.~Xu, D.~W.~K. Ng, and R.~Schober, ``{IRS}-assisted green communication
  systems: {Provable} convergence and robust optimization,'' \emph{{IEEE}
  Trans. Commun.}, vol.~69, no.~9, pp. 6313--6329, Sep. 2021.

\bibitem{pan2020multicell}
C.~Pan, H.~Ren, K.~Wang, W.~Xu, M.~Elkashlan, A.~Nallanathan, and L.~Hanzo,
  ``Multicell {MIMO} communications relying on intelligent reflecting
  surfaces,'' \emph{IEEE Trans. Wireless Commun.}, vol.~19, no.~8, pp.
  5218--5233, Aug. 2020.

\bibitem{wu2019beamforming}
Q.~Wu and R.~Zhang, ``Beamforming optimization for wireless network aided by
  intelligent reflecting surface with discrete phase shifts,'' \emph{{IEEE}
  Trans. Commun.}, vol.~68, no.~3, pp. 1838--1851, Mar. 2020.

\bibitem{liu2021compact}
K.~Liu, Z.~Zhang, L.~Dai, and L.~Hanzo, ``Compact user-specific reconfigurable
  intelligent surfaces for uplink transmission,'' \emph{IEEE Trans. Commun.},
  vol.~70, no.~1, pp. 680--692, Jan. 2022.

\bibitem{ma2021wideband}
S.~Ma, W.~Shen, J.~An, and L.~Hanzo, ``Wideband channel estimation for
  {IRS}-aided systems in the face of beam squint,'' \emph{IEEE Trans. Wireless
  Commun.}, vol.~20, no.~10, pp. 6240--6253, Oct. 2021.

\bibitem{alwazani2020intelligent}
Q.-U.-A. Nadeem, H.~Alwazani, A.~Kammoun, A.~Chaaban, M.~Debbah, and M.-S.
  Alouini, ``Intelligent reflecting surface-assisted multi-user {MISO}
  communication: Channel estimation and beamforming design,'' \emph{IEEE Open
  J. Commun. Soc.}, vol.~1, no.~4, pp. 661--680, May 2020.

\bibitem{mishra2019channel}
D.~Mishra and H.~Johansson, ``Channel estimation and low-complexity beamforming
  design for passive intelligent surface assisted {MISO} wireless energy
  transfer,'' in \emph{Proc. {IEEE} Int. Conf. Acoust., Speech Signal Process.
  (IEEE ICASSP'19)}, Brighton, U.K., May 2019, pp. 4659--4663.

\bibitem{Dai2020reconfigurable}
L.~{Dai}, B.~{Wang}, M.~{Wang}, X.~{Yang}, J.~{Tan}, S.~{Bi}, S.~{Xu},
  F.~{Yang}, Z.~{Chen}, M.~D. {Renzo}, C.~{Chae}, and L.~{Hanzo},
  ``Reconfigurable intelligent surface-based wireless communications: Antenna
  design, prototyping, and experimental results,'' \emph{IEEE Access}, vol.~8,
  pp. 45\,913--45\,923, Mar. 2020.

\bibitem{hu2019two}
C.~Hu, L.~Dai, S.~Han, and X.~Wang, ``Two-timescale channel estimation for
  reconfigurable intelligent surface aided wireless communications,''
  \emph{{IEEE} Trans. Commun.}, vol.~69, no.~11, pp. 7736--7747, Nov. 2021.

\bibitem{glover1989tabu}
F.~Glover, ``Tabu search—part {I},'' \emph{ORSA J. Comput.}, vol.~1, no.~3,
  pp. 190--206, Aug. 1989.

\bibitem{wang2017analysis}
X.~Wang, M.~Amin, and X.~Cao, ``Analysis and design of optimum sparse array
  configurations for adaptive beamforming,'' \emph{{IEEE} Trans. Signal
  Process.}, vol.~66, no.~2, pp. 340--351, Jan. 2018.

\bibitem{rocca2016unconventional}
P.~Rocca, G.~Oliveri, R.~J. Mailloux, and A.~Massa, ``Unconventional phased
  array architectures and design methodologies—{A} review,'' \emph{Proc.
  IEEE}, vol. 104, no.~3, pp. 544--560, Mar. 2016.

\bibitem{liu2018programmable}
F.~{Liu}, A.~{Pitilakis}, M.~S. {Mirmoosa}, O.~{Tsilipakos}, X.~{Wang}, A.~C.
  {Tasolamprou}, S.~{Abadal}, A.~{Cabellos-Aparicio}, E.~{Alarc{\'o}n},
  C.~{Liaskos}, N.~V. {Kantartzis}, M.~{Kafesaki}, E.~N. {Economou}, C.~M.
  {Soukoulis}, and S.~{Tretyakov}, ``Programmable metasurfaces: State of the
  art and prospects,'' in \emph{Proc. {IEEE} Int. Symp. Circuits Syst. (IEEE
  ISCAS'18)}, Florence, Italy, May 2018, pp. 1--5.

\bibitem{tan2021hybrid}
J.~Tan, S.~Suo, and H.~Qin, ``Hybrid precoding codebook design in
  millimetre-wave massive {MIMO} systems with low-resolution phase shifters,''
  \emph{IET Commun.}, pp. 1--15, May 2021.

\bibitem{yang2016A}
H.~{Yang}, F.~{Yang}, S.~{Xu}, Y.~{Mao}, M.~{Li}, X.~{Cao}, and J.~{Gao}, ``A
  1-bit $10 \times 10$ reconfigurable reflectarray antenna: Design,
  optimization, and experiment,'' \emph{IEEE Trans. Antennas Propag.}, vol.~64,
  no.~6, pp. 2246--2254, Jun. 2016.

\bibitem{zhao2022rethinking}
X.~Zhao, S.~Lu, Q.~Shi, and Z.-Q. Luo, ``Rethinking {WMMSE}: {Can} its
  complexity scale linearly with the number of {BS} antennas?'' \emph{arXiv
  preprint arXiv:2205.06225}, May 2022.

\bibitem{rubinstein2013cross}
R.~Y. Rubinstein and D.~P. Kroese, ``The cross-entropy method: A unified
  approach to combinatorial optimization, {Monte-Carlo} simulation and machine
  learning,'' \emph{Springer Science and Business Media}, 2013.

\bibitem{Ozdogan'19}
{Ö. Özdogan}, E.~{Björnson}, and E.~G. {Larsson}, ``Intelligent reflecting
  surfaces: Physics, propagation, and pathloss modeling,'' \emph{IEEE Wireless
  Commun. Lett.}, vol.~9, no.~5, pp. 581--585, May 2020.

\bibitem{zhang2020joint}
Z.~Zhang and L.~Dai, ``A joint precoding framework for wideband reconfigurable
  intelligent surface-aided cell-free network,'' \emph{IEEE Trans. Signal
  Process.}, vol.~69, pp. 4085--4101, Aug. 2021.

\end{thebibliography}

\begin{IEEEbiography}[{\includegraphics[width=1in,height=1.25in,clip,keepaspectratio]{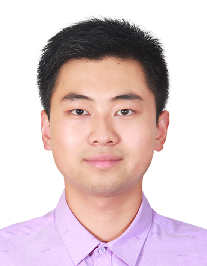}}]{Ruochen Su} received the B.S. degree in the Department of Electronic Engineering, Tsinghua University, Beijing, China, in 2018, where he is currently pursuing his Ph. D. degree. His research interests include reconfigurable intelligent surface (RIS), massive MIMO, and Terahertz communications.
\end{IEEEbiography}

\begin{IEEEbiography}[{\includegraphics[width=1in,height=1.25in,clip,keepaspectratio]{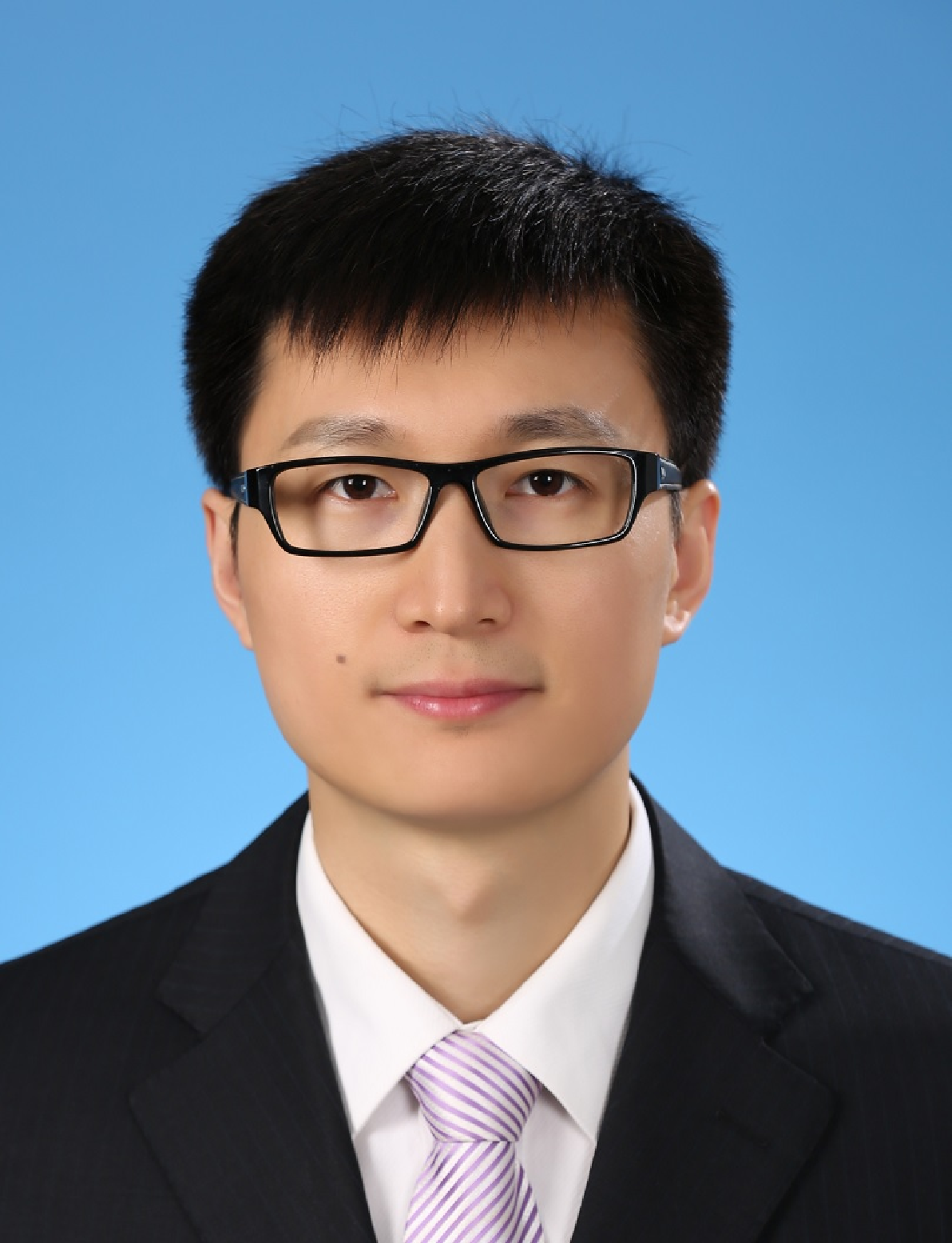}}]{Linglong Dai} received the B.S. degree from Zhejiang University, Hangzhou, China, in 2003, the M.S. degree (with the highest honor) from the China Academy of Telecommunications Technology, Beijing, China, in 2006, and the Ph.D. degree (with the highest honor) from Tsinghua University, Beijing, China, in 2011. From 2011 to 2013, he was a Postdoctoral Research Fellow with the Department of Electronic Engineering, Tsinghua University, where he was an Assistant Professor from 2013 to 2016, an Associate Professor since from 2016 to 2022, and has been a Professor since 2022. His current research interests include massive MIMO, reconfigurable intelligent surface (RIS), millimeter-wave and Terahertz communications, machine learning for wireless communications, and electromagnetic information theory.
	 
	He has authored or coauthored over 80 IEEE journal articles and over 50 IEEE conference papers. He also holds 19 granted patents. He has coauthored the book MmWave Massive MIMO: A Paradigm for 5G (Academic Press, 2016). He has received five IEEE Best Paper Awards at the IEEE ICC 2013, the IEEE ICC 2014, the IEEE ICC 2017, the IEEE VTC 2017-Fall, and the IEEE ICC 2018. He has also received the Tsinghua University Outstanding Ph.D. Graduate Award in 2011, the Beijing Excellent Doctoral Dissertation Award in 2012, the China National Excellent Doctoral Dissertation Nomination Award in 2013, the URSI Young Scientist Award in 2014, the IEEE Transactions on Broadcasting Best Paper Award in 2015, the Electronics Letters Best Paper Award in 2016, the National Natural Science Foundation of China for Outstanding Young Scholars in 2017, the IEEE ComSoc Asia-Pacific Outstanding Young Researcher Award in 2017, the IEEE ComSoc Asia-Pacific Outstanding Paper Award in 2018, the China Communications Best Paper Award in 2019, IEEE Access Best Multimedia Award in 2020, the IEEE Communications Society Leonard G. Abraham Prize in 2020,  and the IEEE ComSoc Stephen O. Rice Prize in 2022, and the IEEE ICC Outstanding Demo Award in 2022. He was listed as a Highly Cited Researcher by Clarivate Analytics from 2020 to 2022.
	
	He is currently serving as an Area Editor of the IEEE Communications Letters. He has also served as an Editor of the IEEE Transactions on Communications (2017-2021), an Editor of the IEEE Transactions on Vehicular Technology (2016-2020), and an Editor of the IEEE Communications Letters (2016-2020). He has also served as a Guest Editor of the IEEE Journal on Selected Areas in Communications, IEEE Journal of Selected Topics in Signal Processing, IEEE Wireless Communications, etc. 
	
	Particularly, he is dedicated to reproducible research and has made a large amount of simulation code publicly available.
\end{IEEEbiography}

\begin{IEEEbiography}[{\includegraphics[width=1in,height=1.25in,clip,keepaspectratio]{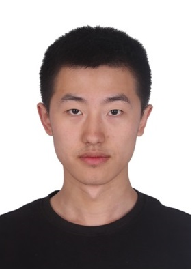}}]{Jingbo Tan} received his B.S. and Ph.D. degree in the Department of Electronic Engineering, Tsinghua University, Beijing, China, in 2017 and 2022, respectively. His research interests include precoding and channel estimation in massive MIMO, THz communications, and reconfigurable intelligent surface aided systems. He has received the IEEE Communications Letters Exemplary Reviewer Award in 2018 and the Honorary Mention in the 2019 IEEE ComSoC Student Competition.
\end{IEEEbiography}

\begin{IEEEbiography}[{\includegraphics[width=1in,height=1.25in,clip,keepaspectratio]{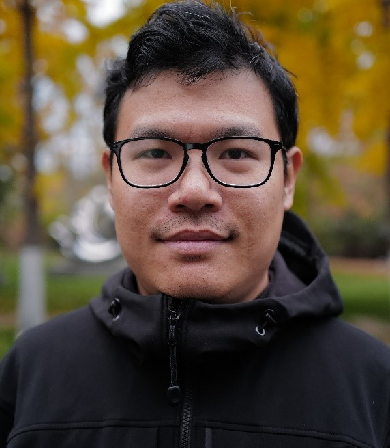}}]{Mo Hao} received the BEng degree from Sichuan University, China in 2006, and the MEng degree from University of Southampton, UK in 2008. He was a senior engineer at Ericsson. He joined the Tsinghua SEM Advanced ICT Laboratory in 2015, where he currently is a researcher. He is in charge of the Beyond 5G related researches and project management. He is participating in the collaboration project on 5G with Tsinghua EE team and also leading the Lab’s independent projects on wireless communication technologies. His research interests include massive MIMO, RIS, and other key technologies for Beyond 5G.
\end{IEEEbiography}

\begin{IEEEbiography}[{\includegraphics[width=1in,height=1.25in,clip,keepaspectratio]{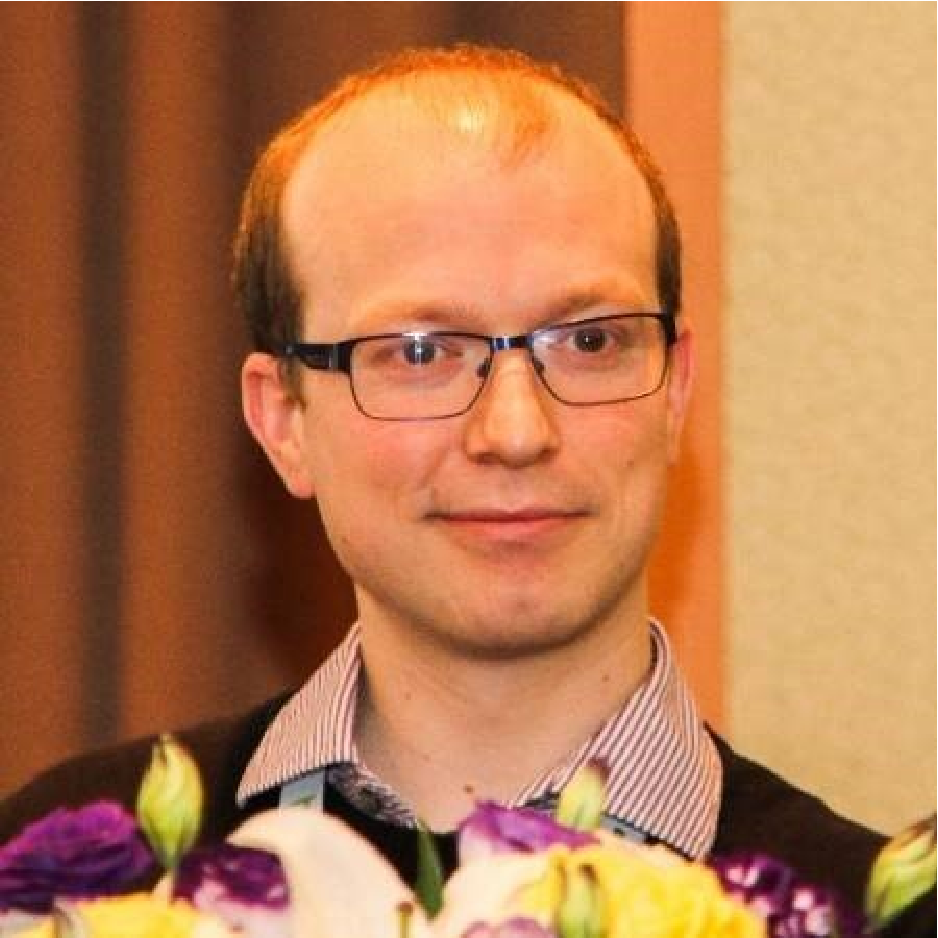}}]{Richard Mackenzie} received the M.Eng. degree in electronic engineering from the University of York and the Ph.D. degree in electronic engineering from the University of Leeds. He has worked as a researcher in wireless communications at BT since 2009 where he is now a Distinguished Engineer. His earlier work covered spectrum management, cognitive radio, femtocells and next-generation technologies. He has been involved in EU projects, in particular QoSMOS, where he was a work package leader for the network architecture design. He was a Spectrum Manager for the London 2012 Olympic and Paralympic Games and was part of BT’s bidding team for spectrum auctions, notably the U.K. 4G spectrum auction in 2013. 
	
	His current research is focussed on creating an ecosystem for RAN virtualization, network automation, and the use of small cells. He is a technical representative for BT at many industry fora, including Telecom Infra Project (TIP), where he is Co-Chair of the ``RAN Intelligence \& Automation'' OpenRAN sub-project; Next Generation Mobile Networks (NGMN), where he was Chair of the “RAN Functional Split and X-Haul” Group; O-RAN Alliance and Small Cell Forum (SCF).
\end{IEEEbiography}

\end{document}